\documentclass[twocolumn, amsmath, amssymb, showpacs, showkeys, aps, pre]{revtex4-1}
\usepackage{graphicx}
\usepackage{dcolumn}
\usepackage{bm}
\usepackage{hyperref}
\usepackage{epstopdf}

\newcommand{\ket}[1]{\left| #1 \right\rangle}
\newcommand{\bra}[1]{\left\langle #1 \right|}

\newcommand{\braket}[2]{\left\langle #1 \right. \left| #2 \right\rangle}

\begin{document}

\title{Path integral Monte Carlo on a lattice: extended states}
\author{Mark O'Callaghan and Bruce N. Miller}
\affiliation{Texas Christian University, Fort Worth, TX 76129}

\date{27 January 2014}

\begin{abstract}
{The equilibrium properties of a single quantum particle (qp) interacting with a classical gas for a wide range of temperatures that explore the system's behavior in the classical as well as in the quantum regime is investigated.  Both the quantum particle and atoms are restricted to the sites of a one-dimensional lattice. A path-integral formalism is developed within the context of the canonical ensemble in which the quantum particle is represented by a closed, variable-step random walk on the lattice. Monte Carlo methods are employed to determine the system's properties. For the case of a free particle, analytical expressions for the energy, its fluctuations, and the qp-qp correlation function are derived and compared with the Monte Carlo simulations. To test the usefulness of the path integral formalism, the Metropolis algorithm is employed to determine the equilibrium properties of the qp for a periodic interaction potential, forcing the qp to occupy extended states. We consider a striped potential in one dimension, where every other lattice site is occupied by an atom with potential $\epsilon$, and every other lattice site is empty. This potential serves as a stress test for the path integral formalism because of its rapid site-to-site variation. An analytical solution was determined in this case by utilizing Bloch's theorem due to the periodicity of the potential. Comparisons of the potential energy, the total energy, the energy fluctuations and the correlation function are made between the results of the Monte Carlo simulations and the analytical calculations.}
\end{abstract}



\maketitle

\section*{ I.  INTRODUCTION}

{The subject matter considered in this work focuses on one-dimensional systems in which a single low-mass particle interacts with more massive atoms and molecules arranged in some configuration. The low-mass particle obeys the laws of Quantum Mechanics whereas the more massive atoms or molecules can be treated classically.  Examples of such low-mass particles are electrons, positrons and positronium. This general problem is important for understanding electron transport in insulating materials, weakly ionized plasmas or any other situation which can be modeled by an excess or solvated electron in a classical gas, liquid or solid. $\left[1\right]$ It also applies to lifetime studies of positrons produced by radioactive sources which have been injected into various materials. $\left[2\right]$ These are examples of an individual quantum particle (qp) interacting with a configuration of more massive classical atoms. In the limit of infinite classical particle mass, we are led to consider the idealized geometry where these particles are fixed in space and both the atoms and quantum particle are restricted to sites on a regular lattice.}

{An important consideration is the interaction time scale, $\tau$, which is available for the measurement process. In the case of the positron or positronium, their inherent lifetimes provide a natural limit for this quantity, where the ortho-positronium lifetime of 140 ns provides an upper bound. $\left[1\right]$ However, there is also a natural time, called the association time, during which the excess electron eventually chemically bonds with the gas atom or molecule. $\left[3\right]$ Thus, for experiments which measure electron currents and positron lifetimes, one must consider whether the classical atom interacting with a quantum particle has sufficient time to respond to the presence of the qp. There are three possible scenarios to consider: (1) the qp is non-thermal; (2) the qp thermalizes but the classical atoms do not respond (quenched case); or (3) the system completely equilibrates (annealed case). Complete self-trapping, in which the environmental classical atoms have time to redistribute themselves in a classical bubble or droplet, apparently occurs in He in the critical region and thus requires case 3. However, disorder-induced (Anderson) localization $\left[4\right]$ is still a possibility in case 2. Even though the thermalization times for positron decay in xenon have been called into question, $\left[5\right]$ at present there is still no clear determination of these times by any known investigation. This study shall only focus on case 2 in the above delineation. That is, a given configuration of atoms in a simulation shall remain unchanged during the simulation. Physically, this means that we are only considering the intermediate - $\tau$ case where there is not sufficient interaction time for the classical atoms in the configuration to react to the presence of the quantum particle. Hence, the interaction goes only one way: the quantum particle "feels" the classical atoms, but not vice versa. In the lattice model considered here, the classical atoms are idealized to have effectively infinite mass since they do not move at all.}

{In this study we shall develop a general approach to solve for the equilibrium properties of such one-dimensional quantum systems for any configuration of atoms. We will then consider the situation where the lattice occupancy is periodic, resulting in extended states of the qp according to Bloch's theorem. This type of periodic atomic configuration is called alternating or striped. }

{The algorithm developed here employs  the tight-binding model to solve the system Schrodinger equation. The tight-binding model is essentially a method to calculate the electronic band structure using an approximate set of wave functions based upon superposition of wave functions for isolated atoms located at each occupied site. The term "tight-binding" refers to the electron being considered tightly bound to the atom in which it is associated such that it has limited interaction with the states of neighboring atoms in the solid. $\left[6\right]$ The tight-binding model has been used extensively in the study of many quantum systems including the study of ultracold atoms on optical lattices, $\left[7\right]$ self-trapping of Bose-Einstein condensates in optical lattices, $\left[8\right]$ Anderson localization, $\left[9\right]$ and the plethora of interesting studies of the behavior of graphene. $\left[10\right]$ }

{We will first consider a free particle using a tight-binding Hamiltonian. The average energy, the average energy fluctuation and the qp-qp correlation function shall be derived analytically. Then, we will focus on the development of a Feynman path-integral approach to solve the same problem of a free quantum particle. We shall also derive expressions for the same aforementioned quantities in the path-integral analogue. This approach for studying the free quantum particle closely follows work performed by Guo and Miller. $\left[11\right]$ However, in studying that work, it was found that there were several errors in the equations which prompted a full re-derivation. This re-derivation is presented first in this work. }

{We have developed a computer algorithm to perform simulations using the Feynman-Kac path-integral, which provides one of the few theoretical methods for exploring the equilibrium properties of a model Hamiltonian directly. In this formalism, a single quantum particle is replaced by a closed chain of say $p$ pseudo-particles, each interacting with the host system through a $p$-reduced potential. The predictions are exact in the limit $p \rightarrow \infty$. $\left[11\right]$ Path integrals are particularly useful for describing the quantum mechanics of an equilibrium system because the canonical distribution for a single particle in the path integral picture becomes isomorphic with that of a classical ring polymer of quasiparticles. $\left[19\right]$ Due to the polymer being in the form of a ring, we have to ensure that the chain closes on itself. In addition, we need to impose boundary conditions on the lattice because of computational limitations.  To more closely approximate an infinite system,  we chose to assert periodic boundary conditions.}

{Here we will show that the form of the path-integral Monte Carlo (PIMC) algorithm used in this work is different than what is seen in continuous systems described by the Schrodinger equation with a given potential. $\left[21\right]$ For a continuous system a free particle, whose Hamiltonian is simply the kinetic energy, has a path-integral whose integrand is just a product of Gaussians, and hence the positions of the pseudo-particles can always be directly sampled. However, it is almost always the case that when a potential is applied, the pseudo-particle positions cannot be directly sampled. The integral has an integrand that is generally not a Gaussian. One then solves the eigenvalue problem for the free particle and constructs a transformation matrix such that the path integral can be described in terms of the normal modes where the kinetic energy Hamiltonian can be directly sampled. The normal modes solution is then transformed back to pseudo-particle positions and the Metropolis algorithm is used to accept or reject the proposed solution based upon the evaluation of the potential energy. The algorithm is used to calculate equilibrium properties such as the energy which can then be used to construct the partition function, correlation functions, etc. }

{We cannot apply the above form of the PIMC algorithm in this problem. The primary reason is that the systems considered in this problem involve a quantum particle interacting with a configuration of classical atoms on a rigid one-dimensional lattice, and the particle positions are confined to be only on lattice sites because the space is assumed discrete. Therefore the standard assumption for the path-integral does not apply. We shall see in the development of this paper that a discrete form of the path integral Monte Carlo algorithm has to be established to account for such confinement in particle positions. In fact, one of the results of this work is the development of such an alternate path-integral Monte Carlo algorithm.}

{Path-integral Monte Carlo algorithms were at first plagued by slow convergence in their earlier existence. But, this issue has been more than compensated by faster computers and clusters which are readily available in most universities and many places of industry. There has also been significant improvements put forward toward faster converging PIMC algorithms both for real-time and imaginary-time approaches. $\left[20\right]$ Hence, path-integral Monte Carlo is an accessible means to perform computations to predict the behavior of quantum systems. The path-integral method has been used to predict the decay rate of the positron $\left[12\right]$ and ortho-positronium $\left[13\right]$ in xenon and to study quantum states of electrons in dense gases. $\left[14\right]$} 

{Our goal is to develop an approach that yields correct equilibrium properties for an arbitrary arrangement of atoms on the periodic lattice. For the model considered here we will first compute the solution to a few equilibrium properties of the free quantum particle on a one-dimensional lattice and then we shall compare those results with the corresponding analytical calculations.  In order to test the efficacy of our method, next we will compare the results for a nontrivial configuration of atoms input into the PIMC program with an analytical solution with the same configuration.  The Schrodinger equation only has known analytical solutions for very few potentials. It is well-known that one can analytically solve the Schrodinger equation for certain periodic configurations using the tight-binding model, and one such model is the alternating or striped configuration considered here. This striped configuration can also be easily implemented in the computer program. It produces a challenging potential for the path integral since it varies as rapidly as possible, once per lattice spacing. Due to the variability of this potential, one typically must consider more and smaller step sizes in the path integral. $\left[15\right]$ Thus, after the discussion of the free quantum particle, an analytical derivation of the solution to the Schrodinger equation with a striped-case potential shall be performed. We shall then derive and calculate the analytical average energy, qp-qp correlation function, and an atom-qp correlation function for the striped case. Finally, we shall compare the results of the developed Monte Carlo program with these striped-case analytical results.}

{The structure of this paper is as follows. In Section II, we will discuss the description of the one-dimensional tight-binding lattice model. Then in Section III we shall derive the analytical solution to the partition function, average energy, the energy fluctuation and the qp-qp correlation function in the canonical ensemble for the free particle. Then, we will derive the path integral analogue of the same corresponding free-particle parameters. We then follow with a derivation of the method to generate the random walks using a conditional probability argument. Next, the results of the PIMC calculations are given and compared with the analytical predictions. Then in Section IV the same approach is followed in the study of the striped case configuration. First, we show analytical derivations for the partition function, energy, average potential energy, the ground state energy and the density matrix for the striped-case problem. We then discuss the Metropolis algorithm and the derivation of the atom-qp correlation function. Next, we compare the PIMC results for the striped-case problem with the corresponding analytical calculations performed using Mathematica. In Section V, we finish the paper by providing a summary and conclusions.}
\section*{ II. DESCRIPTION OF THE MODEL}
{We study a one-dimensional system of a low-mass quantum particle, like an electron or positron, interacting with a rigid one-dimensional lattice partially occupied with more massive atoms.  We suppose that the temperature is sufficiently high that the atoms can be treated classically. The lattice sites are occupied by atoms but some lattice sites are left empty. The space for the quantum particle is also discrete, i.e. the quantum particle only lies on the lattice sites. The qp obeys the Schrodinger equation}
\begin{equation}
\hat{H}\Psi = E\Psi
\end{equation}
{where $\hat{H}$ is the Hamiltonian operator, $E$ is the energy eigenvalue, and $\ket{\Psi}$ is the state vector of the qp.}

{For ease of calculation, it is convenient to employ second quantization. Let $\ket{j}$ denote the state in which the quantum particle is on lattice site $j$. Then the set \{ $\ket{j}$ : $j = \ldots, -L,\ldots, -2, -1, 0, 1, 2, \ldots, L, \ldots$\} forms a basis for our state space. In addition, let $\ket{\hspace{3 pt}}$ denote the vacuum state. We define linear annihilation operators $c_j$ and  $c_j^\dagger$ on the basis vectors as follows:}
\begin{equation}
c_j\ket{\hspace{4 pt}} = 0
\end{equation}
\begin{equation}
 c_j\ket{k} = \delta_{j,k}\ket{\hspace{4 pt}}
\end{equation}
\begin{equation}
c_j^\dagger\ket{\hspace{4 pt}} = \ket{j}
\end{equation}
\begin{equation}
c_j^\dagger\ket{k} = 0
\end{equation}
{We also assert the periodic boundary condition}
\begin{equation}
\Psi_{L + 1} = \Psi_1
\end{equation}
{When the space is changed from continuous to discrete, the differential operators in the Hamiltonian should be changed to difference operators, accordingly. The corresponding Hamiltonian is}
\begin{equation}
\hat{H} = 2t - t\sum_j(c_j^\dagger c_{j + 1} + c_{j + 1}^\dagger c_j) + \sum_j v_j c_j^\dagger c_j
\end{equation}
{where $v_j$  is the potential of the qp on lattice site $j$. In our model, we simply take }
\begin{equation}
v_j = \epsilon n_j
\end{equation}
{where $n_j$ is the number of atoms on lattice site $j$, ($n_j = 0, 1$) and with the choice}
\begin{equation}
t = \frac{\hbar^2} {2ma^2}
\end{equation}
{in which $m$ is the mass of the qp and $a$ is the lattice spacing, we get the discrete approximation of the continuous Hamiltonian. This happens to be the Hamiltonian from the tight-binding model. Without loss of generality we choose units such that $t =  1$ in all of the following.}
\section*{ III. FREE PARTICLE ON THE LATTICE}
{When there is an interaction between the low-mass particle and the atoms, generally the problem cannot be solved analytically. We are going to employ a Monte Carlo calculation by way of a discrete version of the Feynman - Kac path integral to study the system.  In the free particle case there are no atoms at all, $v_j$ = 0 for any j, and we can get the exact analytical solution of the Schrodinger equation. We also carry out Monte Carlo calculations for the free particle case so that we can compare them with the exact analytical solutions. For the interacting case we can almost always only rely on our Monte Carlo calculations.}
\subsection*{ A. Analytical solution}
\subsection*{\textit{ 1. Energy spectrum and eigenstates}}
{For the free particle case, it is easy to see that we have the following solution for the eigenstates of the Schrodinger equation}
\begin{equation}
\Psi_j^{(\alpha)} = \frac{1} {\sqrt{L}} \exp \left(\frac{2 \pi i \alpha j}{L}\right), \alpha = 1, 2, \ldots, L 
\end{equation}
{where $L$ is the lattice size. Substitute this solution into Eq. (1), notice $v_j = 0$, then we have the energy eigenvalues:}
\begin{equation}
E^{(\alpha)} = 2t - 2tcos\left(\frac{2 \pi \alpha} {L}\right), \alpha = 1, 2, \ldots, L
\end{equation}
{This is the energy spectrum. If $L \rightarrow \infty$, we get a single continuous energy band and a two-fold degeneracy because for each energy $E^{(\alpha)}$ there are two different states $\Psi^{(\alpha)}$ and $\Psi^{(L - \alpha)}$ carrying the same energy while having opposite direction of energy flux.}
\subsection*{\textit{ 2. The canonical ensemble}}
{We study the canonical ensemble of such free particle systems at finite temperature. We compute the mean energy of the qp, its mean square fluctuation, and the quantum correlation of the qp with itself along the lattice.}

{The equilibrium density matrix of such an ensemble is $\exp\left(-\beta \hat{H}\right)$ where, as usual, $\beta$ is the inverse temperature in appropriate units. We need to calculate the partition function per lattice site, $Z/L$}
\begin{equation}
\frac{Z}{L} = \frac{1}{L}\sum_{\alpha = 1}^L \exp\left(-\beta E^{(\alpha)}\right)
\end{equation}
{For the free particle, this can be expressed as:}
\begin{equation}
\frac{Z}{L} = \frac{1}{L} e^{-2 \beta t } \sum_{\alpha = 1}^L  e^{2 \beta t cos \left(\frac{2 \pi \alpha}{L}\right)} \nonumber
\end{equation}
{Take the limit as $L \rightarrow$ $\infty$ and make a change of variable to get}
\begin{equation}
\frac{Z}{L} = e^{-2 \beta t } \lim_{L \rightarrow \infty} \left\{  \frac{1}{L} \frac{L}{2 \pi} \sum_{\alpha = \frac{2 \pi}{L}}^{2 \pi} du  e^{2 \beta t cosu}\right\} \nonumber
\end{equation}
{The summation now changes to an integral in $u$.}
\begin{equation}
\frac{Z}{L} = e^{-2 \beta t } \frac{1}{\pi} \int_0^{\pi}du e^{2 \beta t cosu} = e^{-2 \beta t } I_0\left(2 \beta t \right)
\end{equation}
{where $I_0$ is the zeroeth order Modified Bessel function. 

{Modified Bessel functions play a central role in the solutions for the equilibrium properties for the free particle. We introduce the Modified Bessel function at this point and state the important recurrence relations we will need in the derivations to follow. $\left[16\right]$ }

{ First, in general, the $n^{th}$ order Modified Bessel function is given by}
\begin{equation}
I_n\left(z\right) = \frac{1}{\pi}\int_0^{\pi}due^{zcosu}cos\left(nu\right)
\end{equation}
\\
{Other important relations for Modified Bessel functions which we will use in this paper are}
\begin{equation}
I_n^{'}\left(z\right) =  I_{n + 1}\left(z\right) + \frac{n}{z}I_n\left(z\right)
\end{equation}
\begin{equation}
I_n^{'}\left(z\right) = I_{n - 1}\left(z\right) - \frac{n}{z}I_n\left(z\right)
\end{equation}

{The expectation value of the energy is}
\begin{equation}
\langle \hat{H} \rangle = 2t + \langle \hat{H'} \rangle 
\end{equation}
{where $H'$ is the tight-binding Hamiltonian and here the angle brackets represent the thermal average:}
\begin{equation}
\hat{H'} = -t \sum_j \left( c_j^{\dagger} c_{j + 1} + c_{j + 1}^{\dagger} c_j \right)
\end{equation}
{In general, for any operator $\hat{A}$, it is true that}
\begin{equation}
\langle \hat{A} \rangle = \frac{Tr[ \hat{A} e^{-\beta \hat{H}} ]} {Z} \nonumber
\end{equation}
{So,}
\begin{equation}
\langle \hat{H'} \rangle = \frac{\frac{1}{L} Tr[ \hat{H'} e^{-\beta \hat{H}} ]}{\frac{Z}{L}}
\end{equation}
\begin{equation}
\langle \hat{H'} \rangle = \frac{\frac{-1}{L} \frac{\partial}{\partial \beta} Tr[ e^{-\beta \hat{H}} ]}{\left( \frac{Z}{L} \right)} - 2t  \nonumber
\end{equation}
{But, }
\begin{equation}
Tr[ e^{-\beta \hat{H}} ] = \sum_{\alpha} \bra{\alpha} e^{-\beta \hat{H}} \ket{\alpha} = \sum_{\alpha} e^{-\beta E^{(\alpha)}} \nonumber
\end{equation}
{Therefore, we can write}
\begin{equation}
\langle \hat{H'} \rangle = \frac{-\frac{\partial}{\partial \beta}\left( \frac{1}{L} \sum_{\alpha} e^{-\beta E^{(\alpha)}} \right)}{\frac{Z}{L}} - 2t \nonumber
\end{equation}
{Taking the limit as $L \rightarrow \infty$, and thus changing the summation to an integral, we obtain}
\begin{equation}
\langle \hat{H'} \rangle = \frac{\frac{-\partial}{\partial \beta} \left( e^{-2 \beta t} I_0 \left( 2 \beta t \right) \right)}{e^{-2 \beta t} I_0 \left( 2 \beta t \right)} -2t \nonumber
\end{equation} 
{After performing the partial differentiation in the first term and then some algebraic simplification, we get}
\begin{equation}
\langle \hat{H'} \rangle = \frac{-\frac{\partial}{\partial \beta} I_0 \left( 2 \beta t \right)}{I_0 \left( 2 \beta t \right)}
\end{equation}
{Applying Modified Bessel function recurrence relations, we arrive at the following form for the expectation value for the Hamiltonian:}
\begin{equation}
\langle \hat{H} \rangle = 2t - \frac{2t I_1 \left( 2 \beta t \right)}{I_0 \left( 2 \beta t \right)} 
\end{equation}

{Similarly, we can obtain the average energy fluctuation per site}
\begin{equation}
\langle \hat{H'}^2 \rangle - \langle \hat{H'} \rangle^2 \nonumber
\end{equation}
{Since $\hat{H'} = \hat{H} - 2t$  then }
\\
{$\hat{H'}^2 = \hat{H}^2 -4t\hat{H} + 4t^2$ and }
\begin{equation}
\langle \hat{H'}^2 \rangle =\frac{ \frac{1}{L} Tr[ \hat{H'}^2 e^{-\beta \hat{H}}]}{\frac{Z}{L}} \nonumber
\end{equation}
{One can then write}
\begin{equation}
\langle \hat{H'}^2 \rangle = \frac{\frac{1}{L} Tr[ \hat{H}^2 e^{-\beta \hat{H}} ]}{\left( \frac{Z}{L} \right)} - \frac{\frac{4t}{L} Tr[ \hat{H} e^{-\beta \hat{H}} ]}{\left( \frac{Z}{L} \right)}  \nonumber
\end{equation}
\begin{equation}
+ \frac{\frac{4t^2}{L} Tr [e^{-\beta \hat{H}} ]}{\left( \frac{Z}{L} \right)} \nonumber
\end{equation}
{Applying similar mathematical arguments as in the derivation of $\langle \hat{H} \rangle$ above, we obtain}
\begin{equation}
\langle \hat{H'}^2 \rangle = \frac{\frac{\partial^2}{\partial \beta^2}\left( e^{-2 \beta t} I_0 \left( 2 \beta t \right) \right)}{e^{-2 \beta t} I_0 \left( 2 \beta t \right)}  \nonumber
\end{equation}
\begin{equation}
+ \frac{4t \frac{\partial}{\partial \beta} \left( e^{-2 \beta t} I_0 \left( 2 \beta t \right)\right)}{e^{-2 \beta t} I_0 \left( 2 \beta t \right)} + 4t^2 \nonumber
\end{equation}
{With some application of partial differentiation with respect to $\beta$ and some algebraic cancelations and simplifications, we get}
\begin{equation}
\langle \hat{H'}^2 \rangle = \frac{\partial^2 I_0 \left(2 \beta t \right)}{\partial \beta^2}\big/I_0 \left( 2 \beta t \right)
\end{equation} 
{But, we can use Modified Bessel function recurrence relations to finally get}
\begin{equation}
\langle \hat{H'}^2 \rangle - \langle \hat{H'} \rangle ^2 =  2t^2 + 2t^2 \frac{I_2 \left( 2 \beta t \right)}{I_0 \left( 2 \beta t \right)} - 4t^2 \frac{I_1^2 \left( 2 \beta t \right)}{I_0^2 \left( 2 \beta t \right)}
\end{equation}

{Following Guo and Miller $\left[11\right]$ we define the qp-qp correlation function. }
\begin{equation}
G_1 \left( n \right) = \langle \sum_j \Psi_j^* \Psi_{j + n} \rangle
\end{equation}
{Since $\Psi_{j + n}$ is the wave function $\Psi_j$ displaced by $n$ lattice sites, $G_1 \left( n \right)$ provides a measure of the mean spread of the qp along the lattice. We will investigate another correlation called the atom-qp correlation function later.}
\\
{Starting from Eq. (24), let's derive an expression for $G_1 \left( n \right)$ for the free particle in terms of Modified Bessel functions. }
\begin{equation}
G_1 \left( n \right) = \frac{\frac{1}{L} \sum_{\alpha} \bra{\alpha} \left( \sum_j \Psi_j^* \Psi_{j + n} \right) e^{-\beta \hat{H}} \ket{\alpha}}{\left( \frac{Z}{L} \right)} \nonumber
\end{equation}
{We can make use of Eq. (10) now, and substitute the appropriate form of the wavefunction}
\begin{equation}
\begin{split}
G_1 \left( n \right) = \frac{1}{\left( \frac{Z}{L} \right)}\frac{1}{L} \sum_{\alpha} \bra{\alpha} \left( \sum_j \left( \frac{1}{\sqrt{L}} exp \left(\frac{-2 \pi i \alpha j}{L} \right) \right) \right. \\
\left. \times \left( \frac{1}{\sqrt{L}} exp \left(\frac{-2 \pi i \alpha \left( j + n \right)}{L} \right) \right) \right) e^{-\beta \hat{H}} \ket{\alpha} \nonumber
\end{split}
\end{equation}
{which then becomes}
\begin{equation}
G_1 \left( n \right) = \frac{\frac{1}{L^2} \sum_{\alpha} \left( \sum_j  exp \left( \frac{2 \pi i \alpha n}{L} \right) \right) e^{-\beta E^{\left( \alpha \right)}}}{\left( \frac{Z}{L} \right)} \nonumber
\end{equation}
{Since the summand for the $j$-sum does not depend on $j$, we can write}
\begin{equation}
\sum_j exp \left( \frac{2 \pi i \alpha n}{L} \right) = L\text{ } exp \left( \frac{2 \pi i \alpha n}{L} \right) \nonumber
\end{equation}
{Then the qp-qp correlation function becomes}
\begin{equation}
G_1 \left( n \right) = \frac{ \sum_{\alpha} \frac{1}{L} exp \left( -\beta E^{(\alpha)} \right) exp \left( \frac{2 \pi i \alpha n}{L} \right)}{\sum_{\alpha} \frac{1}{L} exp \left( -\beta E^{(\alpha)} \right)}
\end{equation}
{Let $N$ be the numerator of Eq. (25). Then}
\begin{equation}
N = \sum_{\alpha} \frac{1}{L} exp \left( -\beta E^{(\alpha)} \right) exp \left( \frac{2 \pi i \alpha n}{L} \right)  \nonumber
\end{equation}
{From Eq. (11), we can substitute the specific form of the energy and write}
\begin{equation}
N = \frac{1}{L} e^{-2 \beta t} \sum_{\alpha = 1}^L  e^{2 \beta t cos \left( \frac{2 \pi \alpha}{L} \right)} e^{\frac{2 \pi i \alpha n}{L}}  \nonumber
\end{equation}
{Making an appropriate change of variables, taking the limit as $L \rightarrow \infty$, and utilizing the definition of the Modified Bessel function given by Eq. (14) we can write }
\begin{equation}
N = e^{-2 \beta t} \frac{1}{\pi} \int_0^{\pi}  du e^{2 \beta t cosu} cosnu = e^{-2 \beta t} I_n \left( u \right)  \nonumber
\end{equation}
{We have already determined that}
\\
{$\frac{Z}{L} = e^{-2 \beta t} I_0 \left( 2 \beta t \right)$, therefore we can write}
\begin{equation}
G_1 \left( n \right) = \frac{I_n \left( 2 \beta t \right)}{I_0 \left( 2 \beta t \right)}
\end{equation}
{Essentially, we have determined that the qp-qp correlation function, or how the quantum particle spreads itself out on the one-dimensional lattice, is simply the quotient of two different orders of the Modified Bessel function.}
\subsection*{ B. Path integral formalism}
{In general, for an arbitrary interacting system, there is no direct way to compute the quantum trace. Lacking a quantum computer, we have to find some equivalent classical system which we can sample by standard means. For each physical observable we want to investigate, we need to construct a corresponding "classical" operator in the path integral formalism. These transformed operators become the random functions which are averaged in the Monte Carlo calculations.}
\subsection*{\textit{ 1. Partition function}}
{To demonstrate the path integral reformulation, we start with the partition function. In the $\ket{j}$ representation the partition function is}
\begin{equation}
Z = Tr[e^{-\beta \hat{H'}}] = \sum_{j_1} \bra{j_1} e^{-\beta \hat{H'}} \ket{j_1}
\end{equation}
{We split the operator into p factors,}
\begin{equation}
e^{-\beta \hat{H'}} = \left( e^{\frac{-\beta \hat{H'}}{p}} \right)^p = e^{\frac{-\beta \hat{H'}}{p}}e^{\frac{-\beta \hat{H'}}{p}}\cdots e^{\frac{-\beta \hat{H'}}{p}} 
\end{equation}
{and insert the identity operators,}
\begin{equation}
\sum_{j_{\alpha}} \ket{j_{\alpha}} \bra{j_{\alpha}} = 1
\end{equation}
{Making use of the Trotter formula, we have}
\begin{equation}
Z = \sum_{j_1}\sum_{j_2} \cdots \sum_{j_p} \bra{j_1} e^{\frac{-\beta \hat{H'}}{p}} \ket{j_2}\bra{j_2} e^{\frac{-\beta \hat{H'}}{p}} \ket{j_3}  \nonumber
\end{equation}
\begin{equation}
\times \cdots \bra{j_p} e^{\frac{-\beta \hat{H'}}{p}} \ket{j_1} \nonumber
\end{equation}
\begin{equation}
Z = \sum_{j_1}\sum_{j_2} \cdots \sum_{j_p} \prod_{\alpha = 1}^p \bra{j_{\alpha}} e^{\frac{-\beta \hat{H'}}{p}} \ket{j_{\alpha + 1}}
\end{equation}
{where}
\begin{equation}
j_{p + 1} = j_1
\end{equation}
{We calculate the matrix element in Eq. (30) to be}
\begin{equation}
\bra{j} e^{-\frac{\beta \hat{H'}}{p}} \ket{k} = \sum_{\alpha = 1}^L \braket{j}{\alpha}\bra{\alpha}e^{-\frac{\beta \hat{H'}}{p}}\ket{\alpha}\braket{\alpha}{k} \nonumber
\end{equation}
{expanding in terms of the eigenvectors of $\hat{H'}$.}
\\
{We can then make use of the basic relations $\Psi_j^{*} = \braket{j}{\alpha}$ and $\Psi_k = \braket{\alpha}{k}$ and rewrite this last equation as}
\begin{equation}
\bra{j} e^{-\frac{\beta \hat{H'}}{p}} \ket{k} = \sum_{\alpha = 1}^L \Psi_j^* \left( \alpha \right) e^{-\frac{\beta}{p} \left( E^{(\alpha)} - 2t \right)} \Psi_k \left( \alpha \right) \nonumber
\end{equation}
{Using Eqs (10) and (11), we can substitute the appropriate form of the energy and eigenfunction for the free particle and write}
\begin{equation}
\bra{j} e^{-\frac{\beta \hat{H'}}{p}} \ket{k} = \frac{1}{L} \sum_{\alpha = 1}^L  e^{\frac{2 \pi i \alpha (k - j)}{L}} e^{\frac{2 \beta t}{p} cos \left( \frac{2 \pi \alpha}{L} \right)}  \nonumber
\end{equation}
{Then, making a change of variable, taking the limit as $L \rightarrow \infty$ and exploiting the definition of the Modified Bessel function, Eq. (14), we get}
\begin{equation}
\bra{j} e^{-\frac{\beta \hat{H'}}{p}} \ket{k} =  I_{j - k} \left( \frac{2 \beta t}{p} \right) 
\end{equation}
{We finally obtain the partition function as}
\begin{equation}
Z = \sum_{j_1} \sum_{j_2} \cdots \sum_{j_p} \prod_{\alpha = 1}^p I_{j_{\alpha} - j_{\alpha + 1}} \left( \frac{2 \beta t}{p} \right)
\end{equation}
{where the summation is over all $j_1$, $j_2$, $\ldots$, $j_p$. We can identify the sequence $\vec{j}$ = ($j_1$, $j_2$, $\ldots$, $j_p$) with a $p$-step closed random walk on a lattice which starts at $j_1$ and has steps with displacement}
\begin{equation}
s_{\alpha} = j_{\alpha + 1} - j_{\alpha} 
\end{equation} 
{and the factor}
\begin{equation}
\prod_{\alpha = 1}^p I_{j_{\alpha} - j_{\alpha + 1}} \left( \frac{2 \beta t}{p} \right)
\end{equation}
{as the probability (when properly normalized) assigned to each random walk.}
\\
{Notice that a sample walk $j_1$, $j_2$, $\ldots$, $j_p$ is required to satisfy the constraint $j_{p + 1} = j_1$, or}
\begin{equation}
\sum_{\alpha = 1}^p s_{\alpha} = 0
\end{equation}
{so the random walk is closed.}
\\
{We can also interpret the form of $Z$ from another point of view. We define}
\begin{equation}
\Phi \left( \vec{j} \right) = \Phi \left(j_1, j_2, \ldots, j_p \right) \nonumber
\end{equation}
\begin{equation}
= -\frac{1}{\beta} \sum_{\alpha = 1}^p ln \text{ } I_{j_{\alpha} - j_{\alpha + 1}} \left( \frac{2 \beta t}{p} \right)
\end{equation}
{and}
\begin{equation}
\phi \left( j \right) = -\frac{1}{\beta} ln \text{ } I_{j_{\alpha} - j_{\alpha + 1}} \left( \frac{2 \beta t}{p} \right)
\end{equation}
{Then}
\begin{equation}
\Phi \left( \vec{j} \right) = \sum_{\alpha = 1}^p \phi \left( j_{\alpha} - j_{\alpha + 1} \right)
\end{equation}
{and}
\begin{equation}
\prod_{\alpha = 1}^p I_{j_{\alpha} - j_{\alpha + 1}} \left( \frac{2 \beta t}{p} \right) = exp \left( -\beta \Phi \left(j_1, j_2, \ldots, j_p \right) \right) 
\end{equation}
{We can then interpret the probability}
\begin{equation}
\prod_{\alpha = 1}^p I_{j_{\alpha} - j_{\alpha + 1}} \left( \frac{2 \beta t}{p} \right)
\end{equation}
{as the ensemble probability of certain classical systems}
\begin{equation}
exp \left( -\beta \Phi \left(j_1, j_2, \ldots, j_p \right) \right)
\end{equation}
{This classical system takes the form of a closed ring polymer on the lattice consisting of $p$ particles with the interaction energy $\Phi \left(j_1, j_2, \ldots, j_p \right)$. Effectively, each polymer element is only directly coupled to its nearest neighbors in the chain (not necessarily the nearest lattice site) through the interaction $\phi$ which depends on both $\beta$ and the number of lattice sites separating each pair of polymer elements.}

{Our approach will be to construct random walks by generating sequences of positive and negative integers $s_1$, $s_2$, $\ldots$, $s_p$ according to the probability $P(s_1, s_2, \ldots, s_p)$. We will develop appropriate functions which, when averaged over a large set of walks, converge to the thermal mean of specific physical and statistical quantities, e.g., the energy and the correlation function $G_1 \left( n \right)$.} 
\subsection*{\textit{ 2. Energy}}
{We also need to find a way to calculate the average energy by the Monte Carlo method. We treat the energy in a similar manner as the partition function. As usual,}
\begin{equation}
\hat{H} = \hat{T} + \hat{V}
\end{equation}
{where $\hat{T}$ is the kinetic energy operator and $\hat{V}$ is the potential energy operator. In the free particle case,}
\begin{equation}
\hat{V} = 0
\end{equation}
\begin{equation}
\hat{H'} = \hat{T'} = -t \sum_j \left( c_j^{\dagger} c_{j + 1} + c_{j + 1}^{\dagger} c_j \right)
\end{equation}
{Using the second quantization equations, Eqs (2) - (5), the matrix element in the $\ket{j}$ representation is seen to be}
\begin{equation}
\bra{j_1}\hat{T'} \ket{k} = -t \left(\delta_{j_1, k - 1} + \delta_{j_1, k + 1} \right)
\end{equation}
{Let's now find the  canonical average of $\hat{T'}$}
\begin{equation}
\langle \hat{T'} \rangle = \frac{Tr[ \hat{T'} e^{-\beta \hat{H'}}]}{Z} 
\end{equation}
{Let's first find $Tr[\hat{T'} e^{-\beta \hat{H'}}]$.}
\begin{equation}
Tr[\hat{T'}e^{-\beta \hat{H'}}] = \sum_k \bra{k} \hat{T'} e^{-\beta \hat{H'}} \ket{k} \nonumber
\end{equation}
\begin{equation}
= \sum_k \bra{k} \hat{T'} \left( e^{-\frac{\beta \hat{H'}}{p}} \right) ^p \ket{k} \nonumber
\end{equation}
\begin{equation}
= \sum_k \sum_{j_1} \sum_{j_2} \cdots \sum_{j_p} \bra{k} \hat{T'} \ket{j_1} \bra{j_1} e^{-\frac{\beta \hat{H'}}{p}} \ket{j_2}\bra{j_2} e^{-\frac{\beta \hat{H'}}{p}} \ket{j_3} \nonumber
\end{equation} 
\begin{equation}
\times \cdots \bra{j_p} e^{-\frac{\beta \hat{H'}}{p}} \ket{k} \nonumber
\end{equation}
{Using Eq. (46), we can write}
\begin{equation}
Tr[\hat{T'}e^{-\beta \hat{H'}}] = \sum_{j_1} \sum_{j_2} \cdots \sum_{j_p} -t \left(\delta_{k, j_1 - 1} + \delta_{k, j_1 + 1} \right)  \nonumber
\end{equation}
\begin{equation}
\times \bra{j_1} e^{-\frac{\beta \hat{H'}}{p}} \ket{j_2}\bra{j_2} e^{-\frac{\beta \hat{H'}}{p}} \ket{j_3}\cdots \bra{j_p} e^{-\frac{\beta \hat{H'}}{p}} \ket{k} \nonumber
\end{equation}
{After some algebraic manipulation, we arrive at the following expression for this trace:}
\begin{equation}
Tr[\hat{T'}e^{-\beta \hat{H'}}] =  \nonumber
\end{equation}
\begin{equation}
 \sum_{j_1} \sum_{j_2} \cdots \sum_{j_p}-t \left(\frac{\bra{j_p} e^{-\frac{\beta \hat{H'}}{p}} \ket{j_1 - 1} + \bra{j_p} e^{-\frac{\beta \hat{H'}}{p}} \ket{j_1 + 1}}{\bra{j_p} e^{-\frac{\beta \hat{H'}}{p}} \ket{j_1}} \right)  \nonumber
\end{equation}
\begin{equation}
\times \prod_{\alpha = 1}^p \bra{j_{\alpha}}e^{-\frac{\beta\hat{H'}}{p}}\ket{j_{\alpha + 1}} \nonumber
\end{equation}
{This calculation specifically just evaluated $Tr[\hat{T'}e^{-\beta \hat{H'}}]$ for $j_1$. But, there is nothing special about $j_1$. All $j_1$, $j_2$, \ldots, $j_p$ need to be evaluated. So, there are $p - 1$ more sums just like the one above, one for each $j_{\alpha}$, $\alpha \in [1, p]$. The same argument can also be made for the $j_p$ above, so in general we can write}
\begin{equation}
Tr[\hat{T'}e^{-\beta \hat{H'}}] = \sum_{j_1} \sum_{j_2} \cdots \sum_{j_p}  \left(\frac{-t}{p} F_1 \right) \nonumber
\end{equation}
\begin{equation}
\times \prod_{\alpha = 1}^p \bra{j_{\alpha}}  e^{-\frac{\beta \hat{H'}}{p}} \ket{j_{\alpha + 1}} \nonumber
\end{equation}
{where}
\begin{equation}
F_1 = \sum_{\alpha = 1}^p \frac{\left( \bra{j_{\alpha}} e^{-\frac{\beta \hat{H'}}{p}} \ket{j_{\alpha + 1} - 1} + \bra{j_{\alpha}} e^{-\frac{\beta \hat{H'}}{p}} \ket{j_{\alpha + 1} + 1}\right)}{\bra{j_{\alpha}} e^{-\frac{\beta \hat{H'}}{p}} \ket{j_{\alpha + 1}}} \nonumber
\end{equation}

{The division by $p$ is because we are calculating a sum over $p$ steps and we want to compute the average.}
\\
{Using Eq. (32), we can now express this kinetic energy trace in terms of Modified Bessel functions}
\begin{equation}
Tr[\hat{T'}e^{-\beta \hat{H'}}] = \sum_{j_1} \sum_{j_2} \cdots \sum_{j_p} \left(-\frac{t}{p} F_2 \right) \prod_{\alpha = 1}^p  I_{j_{\alpha} -  j_{\alpha + 1}}  \nonumber
\end{equation}
{where}
\begin{equation}
F_2 = \sum_{\alpha = 1}^p \frac{ I_{\left( j_{\alpha} -  j_{\alpha + 1} \right) + 1} +  I_{\left( j_{\alpha} -  j_{\alpha + 1} \right) - 1 }}{ I_{j_{\alpha} -  j_{\alpha + 1}}} \nonumber
\end{equation}
{We can now make use of a Modified Bessel function recurrence relation to write}
\begin{equation}
Tr[\hat{T'}e^{-\beta \hat{H'}}] = \sum_{j_1} \sum_{j_2} \cdots \sum_{j_p} -\frac{2t}{p}\left(\sum_{\alpha = 1}^p \frac{ I'_{ j_{\alpha} -  j_{\alpha + 1}}\left( \frac{2 \beta t}{p} \right)}{ I_{j_{\alpha} -  j_{\alpha + 1}}\left( \frac{2 \beta t}{p} \right)}\right) \nonumber
\end{equation}
\begin{equation}
\times \prod_{\alpha = 1}^p  I_{j_{\alpha} -  j_{\alpha + 1}}\left( \frac{2 \beta t}{p} \right)
\end{equation}
{where the prime denotes differentiation in $\beta$.}
\\
{We can see from Eq. (48) that the average of a physical observable $\hat{\Theta}$ has the form}
\begin{equation}
\sum_{\text{walks}} \Theta_{\text{cl}} \left( \text{walk} \right) \times \text{Probability} \left( \text{walk} \right)
\end{equation}
{where $\hat{\Theta}_{cl}$ is a function defined on a walk and is the counterpart of a quantum operator $\hat{\Theta}$ in the classical system isomorphism. Each quantum operator has a corresponding classical operator in the isomorphic ensemble of polymer systems. Thus, the quantum kinetic energy operator is}
\begin{equation}
\hat{T} = 2t - t \sum_j \left( c_j^{\dagger} c_{j + 1}  + c_{j + 1}^{\dagger} c_j \right)
\end{equation}
{and its classical analogue is}
\begin{equation}
\hat{\tau} = 2t - \frac{2t}{p} \sum_{\alpha = 1}^p \frac{ I'_{ j_{\alpha} -  j_{\alpha + 1}}\left( \frac{2 \beta t}{p} \right)}{ I_{j_{\alpha} -  j_{\alpha + 1}}\left( \frac{2 \beta t}{p} \right)}
\end{equation}
\subsection*{\textit{ 3. Energy fluctuation}}
{We will treat the square of the energy similarly in order to compute the energy fluctuation. First we calculate the matrix element $\bra{j_1} \hat{T'^2} \ket{k}$. We begin by writing}
\begin{equation}
\bra{j_1} \hat{T'^2} \ket{k} = t^2 \bra{j_1} \left(\sum_j \left( c_j^{\dagger} c_{j + 1} + c_{j + 1}^{\dagger} c_j \right) \right)  \nonumber
\end{equation}
\begin{equation}
\times \left( \sum_l \left( c_l^{\dagger} c_{l + 1} + c_{l + 1}^{\dagger} c_l \right)\right) \ket{k} \nonumber
\end{equation}
{After utilizing the second quantization relations, Eqs (2) - (5), we find this matrix element to be}
\begin{equation}
\bra{j_1} \hat{T'^2} \ket{k} =  t^2 \left( \delta_{j_1, k - 2} + 2 \delta_{j_1, k} +  \delta_{j_1, k + 2} \right)
\end{equation}
{Employing the trace $Tr[ \hat{T'^2} e^{-\beta \hat{H'}}]$ and using the Trotter formula in combination with Eq. (52), we can write}
\begin{equation}
Tr[ \hat{T'^2} e^{-\beta \hat{H'}}] =   \nonumber
\end{equation}
\begin{equation}
 \sum_{j_1} \sum_{j_2} \cdots \sum_{j_p} t^2 \left( \delta_{j_1, k - 2} + 2 \delta_{j_1, k} +  \delta_{j_1, k + 2} \right) \nonumber
\end{equation}
\begin{equation}
\times \bra{j_1} e^{-\frac{\beta \hat{H'}}{p}} \ket{j_2} \cdots \bra{j_p} e^{-\frac{\beta \hat{H'}}{p}} \ket{k} \nonumber
\end{equation}
{At this point, we then apply exactly the same mathematical arguments that were utilized to attain the corresponding expression for the energy in the last section. After following through these same steps we reach the following equation:}
\begin{equation}
 Tr[ \hat{T'^2} e^{-\beta \hat{H'}}] = \sum_{j_1} \sum_{j_2} \cdots \sum_{j_p} \left( \frac{t^2}{p} \left(F_3 \right) \right) \nonumber
\end{equation}
\begin{equation}
\times \prod_{\alpha = 1}^p \bra{j_{\alpha}} e^{-\frac{\beta \hat{H'}}{p}} \ket{j_{\alpha + 1}} \nonumber
\end{equation}
{where}
\begin{equation}
F_3 =  \sum_{\alpha = 1}^p \frac{ \bra{j_{\alpha}} e^{-\frac{\beta \hat{H'}}{p}} \ket{j_{\alpha} - 2}}{\bra{j_{\alpha}} e^{-\frac{\beta \hat{H'}}{p}} \ket{j_{\alpha + 1}}} \nonumber
\end{equation} 
\begin{equation}
+  \sum_{\alpha = 1}^p \frac{ 2 \bra{j_{\alpha}} e^{-\frac{\beta \hat{H'}}{p}} \ket{j_{\alpha}}}{\bra{j_{\alpha}} e^{-\frac{\beta \hat{H'}}{p}} \ket{j_{\alpha + 1}}} \nonumber
\end{equation}
\begin{equation}
+  \sum_{\alpha = 1}^p \frac{ \bra{j_{\alpha}} e^{-\frac{\beta \hat{H'}}{p}} \ket{j_{\alpha} + 2}}{\bra{j_{\alpha}} e^{-\frac{\beta \hat{H'}}{p}} \ket{j_{\alpha + 1}}} \nonumber
\end{equation}
{Using Eq. (32), we can write this trace in terms of Modified Bessel functions}
\begin{equation}
 Tr[ \hat{T'^2} e^{-\beta \hat{H'}}] = \sum_{j_1} \sum_{j_2} \cdots \sum_{j_p}  \frac{t^2}{p} \left( F_4 \right) \prod_{\alpha = 1}^p I_{j_{\alpha} - j_{\alpha + 1}}  \nonumber
\end{equation} 
{where}
\begin{equation}
F_4 =  \sum_{\alpha = 1}^p \frac{I_{\left(j_{\alpha} - j_{\alpha + 1}\right) + 2} + I_{\left(j_{\alpha} - j_{\alpha + 1}\right)} + I_{\left(j_{\alpha} - j_{\alpha + 1}\right) - 2}}{I_{\left(j_{\alpha} - j_{\alpha + 1}\right)}} \nonumber
\end{equation}
{Using the same recurrence relation in the last section, taking another derivative on it, and then applying that result here, we get}
\begin{equation}
 Tr[ \hat{T'^2} e^{-\beta \hat{H'}}] = \sum_{j_1} \sum_{j_2} \cdots \sum_{j_p}  \frac{4 t^2}{p} \left( \sum_{\alpha = 1}^p \frac{I''_{j_{\alpha} - j_{\alpha + 1}} \left( \frac{2 \beta t}{p} \right)}{I_{j_{\alpha} - j_{\alpha + 1}} \left( \frac{2 \beta t}{p} \right)}\right)  \nonumber 
\end{equation}
\begin{equation}
\times \prod_{\alpha = 1}^p I_{j_{\alpha} - j_{\alpha + 1}} \left( \frac{2 \beta t}{p} \right)
\end{equation}
{The classical isomorphism of $\hat{T'^2}$ can be reached by recalling that $\hat{T'} = \hat{H'}$ and then writing $\langle \hat{H}^2 \rangle$ as a polynomial of the ensemble average of powers of $\langle \hat{H'}\rangle$ as follows :}
\begin{equation}
\langle \hat{H}^2 \rangle = \langle \hat{H'}^2 \rangle +4t\langle \hat{H'} \rangle + 4t^2 \nonumber
\end{equation}
{which we shall denote as $\tau_2$. Therefore, we can write the classical isomorphism of $\hat{T'^2}$ as}
\begin{equation}
\tau_2 = \frac{4t^2}{p} \sum_{\alpha = 1}^p \frac{I''_{j_{\alpha} - j_{\alpha + 1}} \left( \frac{2 \beta t}{p} \right)}{I_{j_{\alpha} - j_{\alpha + 1}} \left( \frac{2 \beta t}{p} \right)} -  \frac{8t^2}{p} \sum_{\alpha = 1}^p \frac{I'_{j_{\alpha} - j_{\alpha + 1}} \left( \frac{2 \beta t}{p} \right)}{I_{j_{\alpha} - j_{\alpha + 1}} \left( \frac{2 \beta t}{p} \right)}  + 4t^2
\end{equation}
\subsection*{\textit{ 4. Correlation function}}
{We are going to compute the correlation function for the free particle using the path integral formalism. From the definition given by Eq. (26), we reformulate it in the occupation number representation to take the form}
\begin{equation}
\hat{G_1} \left( n \right) = \sum_j c_j^{\dagger} c_{j + n}
\end{equation}
{The matrix element that we need is}
\begin{equation}
\bra{j_1} \hat{G_1} \ket{k} = \bra{j_1} \sum_j c_j^{\dagger} c_{j + n} \ket{k} \nonumber    
\end{equation}
\begin{equation}
= \bra{j_1} c_j^{\dagger} \delta_{j + n, k} \ket{\:} \nonumber
\end{equation}
\begin{equation}
= \bra{j_1} c_{k - n}^{\dagger} \ket{\:} \nonumber
\end{equation}
\begin{equation}
= \braket{j_1}{k - n} \nonumber
\end{equation}
\begin{equation}
= \delta_{j_1, k - n}
\end{equation}
{Using Eq. (56) and stepping through the same mathematical process as in the previous two sections, we arrive at the following form for $Tr[\hat{G_1} e^{-\beta \hat{H'}}]$.} 
\begin{equation}
Tr[\hat{G_1} e^{-\beta \hat{H'}}] =  \nonumber
\end{equation}
\begin{equation}
\sum_{j_1} \sum_{j_2} \cdots \sum_{j_p} \left( \frac{1}{p} \sum_{\alpha = 1}^p \frac{I_{j_{\alpha} - j_{\alpha + 1} - n} \left( \frac{2 \beta t}{p}\right)}{I_{j_{\alpha} - j_{\alpha + 1}} \left( \frac{2 \beta t}{p}\right)}\right) \nonumber
\end{equation}
\begin{equation}
\times \prod_{\alpha = 1}^p I_{j_{\alpha} - j_{\alpha + 1}} \left( \frac{2 \beta t}{p}\right)
\end{equation}
{It follows that the classical analogue of the qp-qp correlation function is}
\begin{equation}
\Gamma_1\left( n \right) =  \frac{1}{p} \sum_{\alpha = 1}^p \frac{I_{j_{\alpha} - j_{\alpha + 1} - n} \left( \frac{2 \beta t}{p}\right)}{I_{j_{\alpha} - j_{\alpha + 1}} \left( \frac{2 \beta t}{p}\right)}
\end{equation}
\subsection*{\textit {5. Generation of the random walks}}
{Here we develop a Levy method for sampling the walks. Suppose P($\vec{s}$) is the probability of the sequence of steps $(s_1, s_2, \ldots, s_p)$. Then}
\begin{equation}
P \left( s_1, s_2, \ldots, s_p \right) = C \prod_{\alpha = 1}^p I_{s_{\alpha}} \left( \frac{2 \beta t}{p} \right) \Delta \left( \sum_{\alpha} s_{\alpha} \right)
\end{equation}
{where}
\begin{equation}
\label{eq:mdiv}
\Delta\left( x \right) =
\begin{cases}
1 & \text{x = 0} \\
0 & \text{x $\neq$ 0}
\end{cases}
\end{equation}
{and $C$ is a normalization factor. The presence of a function $\Delta$ arises from the constraint $\sum_{\alpha = 1}^p s_{\alpha} = 0$, which requires the random walk to be closed. Using the Fourier representation of $\Delta$(x)}
\begin{equation}
\Delta \left( x \right) = \frac{1}{2 \pi} \int_{-\pi}^{\pi} e^{i k x } dk
\end{equation}
{and the identity}
\begin{equation}
\sum_{s = -\infty}^{\infty} I_s \left( x \right) cos ks \equiv e^{x cosk} 
\end{equation}
{we find $C$ to be}
\begin{equation}
\sum_{s_1}\sum_{s_2}\cdots\sum_{s_p} P \left( s_1, s_2, \ldots, s_p \right) = 1 \nonumber
\end{equation}
\begin{equation}
C\sum_{s_1}\sum_{s_2}\cdots\sum_{s_p}  \prod_{\alpha = 1}^p I_{s_{\alpha}} \left( \frac{2 \beta t}{p} \right) \Delta \left( \sum_{\alpha} s_{\alpha} \right) = 1  \nonumber
\end{equation}
\begin{equation}
C\sum_{s_1}\sum_{s_2}\cdots\sum_{s_p}  \prod_{\alpha = 1}^p I_{s_{\alpha}} \left( \frac{2 \beta t}{p} \right) \frac{1}{2 \pi} \int_{-\pi}^{\pi} dx e^{i x \sum_{\alpha} s_{\alpha} } = 1  \nonumber
\end{equation}
{Carrying out this calculation, we arrive at a normalization constant that is just 1 divided by the zeroeth order Modified Bessel function.}
\begin{equation}
C = \frac{1}{I_0 \left(2 \beta t \right)}
\end{equation}
{We would like to generate the sequence of numbers $j_1, j_2, \ldots, j_p$ one after another instead of getting them all at once as a group. We ask, given the first $\nu$ integers in the sequence $s_1, s_2, \ldots, s_{\nu}$, what is the conditional probability, $P\left(s_{\nu + 1} | s_1, s_2, \ldots, s_{\nu}\right)$, of getting $s_{\nu + 1}$ next?} 
\\
{As usual, the conditional probability can be expressed in terms of the joint probability as}
\begin{equation}
P\left(s_{\nu + 1} | s_1, s_2, \ldots, s_{\nu}\right) = \frac{P\left( s_1, s_2, \ldots, s_{\nu + 1}\right)}{P\left( s_1, s_2, \ldots, s_{\nu}\right)}
\end{equation}
{where}
\begin{equation}
P\left( s_1, s_2, \ldots, s_{\nu}\right) = \nonumber 
\end{equation}
\begin{equation}
C \sum_{s_{\nu + 1}}\sum_{s_{\nu + 2}}\cdots \sum_{s_p} \left( \prod_{\alpha = 1}^{\nu} I_{s_{\alpha}}\left( \frac{2 \beta t}{p} \right)\right) \nonumber
\end{equation}
\begin{equation}
\times \left( \prod_{\alpha = \nu + 1}^p I_{s_{\alpha}}\left( \frac{2 \beta t}{p} \right)\right) \Delta\left( \sum_{\alpha = 1}^{\nu} s_{\alpha} + \sum_{\alpha = \nu + 1}^p s_{\alpha} \right)
\end{equation}
{is the joint probability for $s_1, s_2, \ldots s_{\nu}$ and again $\Delta$ insures the closure of each random walk.}

{Equation (59) computes the probability of a specific sequence $s_1, s_2, ..., s_p$ in a $p$-step random walk out of all of the possible $p$-step random walks. But, Eq. (65) states given a sequence $s_1, s_2, \ldots s_{\nu} \text{  } (\nu \le p)$ what is the probability of obtaining such a sequence? To calculate this, one needs to consider all of the possible sequences $s_{\nu + 1}, s_{\nu + 2}, \ldots, s_p$. Essentially, we have one particular sequence $s_1, s_2, \ldots s_{\nu}$ set and in order to determine the probability of getting the next $p - \nu$ steps, one must sum over all of these other possibilities for sequences $s_{\nu + 1}, s_{\nu + 2}, \ldots s_p$.}

{In Eq. (65) the first product, $\prod_{\alpha = 1}^{\nu} I_{s_{\alpha}} \left( \frac{2 \beta t}{p}\right)$, does not take part in the multiple summation and can actually be pulled out in front of it. In the same manner, the first summation in the $\Delta$ function does not take part in the multiple summation either and can be considered a constant with respect to the multiple summation.}

{Let}
\begin{equation}
t_{\nu} = \sum_{\alpha = 1}^{\nu} s_{\alpha}
\end{equation}
{be the displacement after $\nu$ steps. Then, as we did earlier by employing the Fourier representation of $\Delta$, we have}
\begin{equation}
P\left( s_1, s_2, \ldots, s_{\nu}\right) = C \left[ \prod_{\alpha = 1}^{\nu} I_{s_{\alpha}} \left( \frac{2 \beta t}{p} \right) \right] \nonumber
\end{equation}
\begin{equation}
\times \sum_{s_{\nu + 1}}\cdots \sum_{s_p}\left[ \prod_{\alpha = \nu + 1}^p I_{s_{\alpha}} \left( \frac{2 \beta t}{p} \right) \right] \Delta \left( t_{\nu} + \sum_{\alpha = \nu + 1}^p s_{\alpha}\right) \nonumber
\end{equation}
{and then immediately we can write}
\begin{equation}
P\left( s_1, s_2, \ldots, s_{\nu}\right) = C \left[ \prod_{\alpha = 1}^{\nu} I_{s_{\alpha}} \left( \frac{2 \beta t}{p} \right) \right]  \nonumber
\end{equation}
\begin{equation}
\times \sum_{s_{\nu + 1}}\cdots \sum_{s_p}\left[ \prod_{\alpha = \nu + 1}^p I_{s_{\alpha}} \left( \frac{2 \beta t}{p} \right) \right]  \nonumber
\end{equation}
\begin{equation}
\times \frac{1}{2 \pi} \int_{-\pi}^{\pi} dx e^{i x \left( t_{\nu} + \sum_{\alpha = \nu + 1}^p s_{\alpha}\right)} \nonumber
\end{equation}
{Carrying through this calculation, we obtain}
\begin{equation}
P\left( s_1, s_2, \ldots, s_{\nu}\right) = C \left[ \prod_{\alpha = 1}^{\nu} I_{s_{\alpha}} \left( \frac{2 \beta t}{p} \right) \right] I_{t_{\nu}} \left( \frac{p - \nu}{p} 2 \beta t \right)
\end{equation}
{Finally, we obtain the conditional probability for the next step in a $p$-step walk.}
\begin{equation}
 P\left( s_{\nu + 1} | s_1, s_2, \ldots, s_{\nu}\right) = \frac{I_{s_{\nu + 1}}\left(\frac{2 \beta t}{p} \right) I_{t_{\nu + 1}}\left(\frac{p - \nu - 1}{p} 2 \beta t \right)}{I_{t_{\nu}}\left(\frac{p - \nu }{p} 2 \beta t \right)}
\end{equation}
\subsection*{ C. Results from the Monte Carlo calculation}
{We generated random walks step by step by partitioning the unit interval by the sequence of subintervals $P\left(0| s_1, s_2, \ldots s_{\nu} \right), P\left(1| s_1, s_2, \ldots s_{\nu} \right), P\left(-1| s_1, s_2, \ldots s_{\nu} \right), $ $P\left(2| s_1, s_2, \ldots s_{\nu} \right), P\left(-2| s_1, s_2, \ldots s_{\nu} \right)$, .... We select a random number from the uniform distribution and determine which interval of the partition it occupies. Then the displacement of the random walk in this step is determined. We add up the contributions for the classical analogue of the operator for the complete random walk and finally we take the mean over all walks to get the average.}

{Using a computer cluster, we calculated the energy and correlation functions over a variety of temperatures and they agreed very well with the analytical results. Figure 1 shows the energy over a range of temperatures. The solid curve is the analytical result, and the crosses show the Monte Carlo results. In Figure 2 we plot the qp-qp correlation function vs. separation $n$ for $\beta$ = 10. The agreement of the Monte Carlo simulations with the analytical results was within $\sim20\%$ up to a spacing of $n = 5$ and then there was disagreement beyond this point due to the occurrence of rare events. Calculations over a wide range of temperatures (not shown) confirms our intuition that when the temperature is lowered the correlation spreads out. It is important to note that for a free particle the discrete path integral is exact for finite Trotter number $p$. In practice, choosing values between 10 and 100, we obtained excellent convergence (see Fig. 1) by sampling $10^6$ independent walks. Typical differences from the exact result were in the fourth decimal place for the energy.    } 
\begin{figure} [h]
\begin{center}
\includegraphics[scale = 0.7]{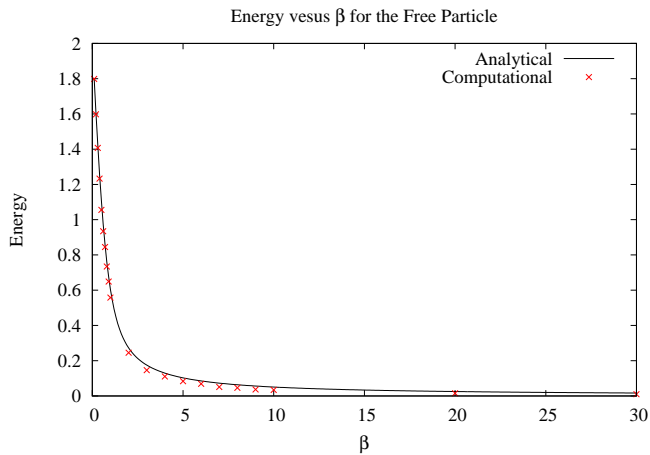} 
\caption{\label {fig:qm/complexfunctions} Energy versus inverse temperature, $\beta$, for a free particle moving on the lattice. The solid curve is a plot of the exact theoretical results. The crosses represent the Monte Carlo simulations. The agreement is outstanding. Error bars are too small to be seen on this scale. }
\end{center}
\end{figure}
\begin{figure} [h]
\begin{center}
\includegraphics[scale=0.6] {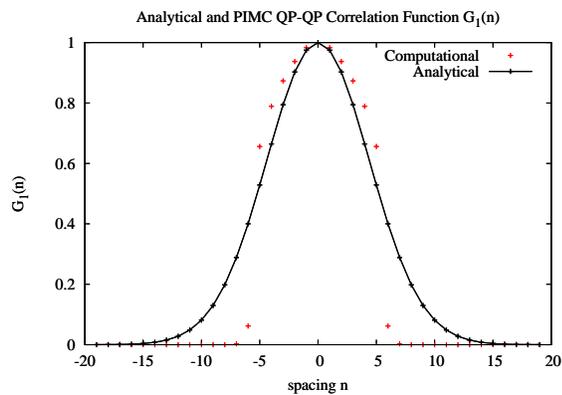} 
\caption{\label {fig:qm/complexfunctions} Self-correlation function, $G_1\left( n \right)$, of the free quantum particle on the lattice for $\beta$ = 10. The solid curve is a plot of the exact theoretical results while the crosses represent the Monte Carlo simulations. Note the excellent agreement except near the shoulders, where the number of significant events becomes small. }
\end{center}
\end{figure}
\section*{ IV. PRESENCE OF ATOMS ON THE LATTICE IN A FIXED PERIODIC CONFIGURATION}
\subsection*{A. Fixed periodic configuration of atoms}
{Now let's consider the interaction between the qp and a set configuration of atoms. This is easiest when the atoms are rigidly fixed on the lattice, then the qp just sees the atoms as a source of an external potential field. This corresponds to a one-way interaction between the qp and the atoms on the lattice, where the qp knows about the atoms, but the atoms do not know about the qp. This type of scenario is commonly referred to as a quenched system.}

{In this paper, we will particularly consider the case of an alternating potential, i.e. the case where every other site on the one-dimensional lattice is occupied by an atom and every other lattice site is empty. For each occupied site, it can only be occupied by at most one atom and this atom is assumed to possess an on-site potential $\epsilon$. This configuration is also known as the "striped" case. Configurations set on a lattice in this manner are known as examples of the tight-binding model. Most potential problems in Quantum Mechanics cannot be solved analytically. There are a handful of problems which have an analytical solution, which include the central potential, the harmonic oscillator potential, the infinite square well potential and very few others. To ensure that the Monte Carlo code being developed is correct, it is desired to be able to establish a configuration of atoms on the lattice and compare the computational solution with an analytical solution. The striped case configuration is one that can be solved analytically and also easily implemented into the PIMC computer program. The following sections of this paper shall provide an analytical derivation of the solution of the Schrodinger equation for the striped case, and also the derivation of the important physical parameters of interest, including the average energy, the energy fluctuation, and the atom-qp correlation function for this striped configuration. }
\subsection*{\textit{ 1. Analytical solution for the configuration of alternating atomic occupation, also known as the striped case}}
{The configuration with alternating atomic occupation, or the striped case, can be described most generally by considering every other lattice site to have potential $a$ and the other lattice sites to have potential $b$, in an arrangement $a$ - $b$ - $a$ - $b$ - $a$ - $b$ - ... - $a$ - $b$. The following tridiagonal matrix describes the solutions to the Schrodinger equation,}

\begin{center}
\[ \textbf{M} = \left( \begin{array}{cccccccccc}
a & -1 & 0 & 0 & 0 & 0 & 0 & 0 & 0 & -1\\
-1 & b & -1 & 0 & 0 & 0 & 0 & 0 & 0 & 0\\
0 & -1 & a & -1 & 0 & 0 & 0 & 0 & 0 & 0\\
\vdots \\
\vdots \\
0 & 0 & 0 & 0 & 0 & 0 & 0 & -1 & a & -1 \\
0 & 0 & 0 & 0 & 0 & 0 & 0 & 0 & -1 & b \end{array} \right) \]
\end{center}
{where the Schrodinger equation is}
\begin{equation}
M\Psi = E\Psi
\end{equation}
{Two sets of difference equations arise from the application of the Schrodinger equation of this form, and they are}
\begin{equation} 
-\left(\Psi_{j - 1} + \Psi_{j + 1}\right) + a\Psi_j = E\Psi_j
\end{equation}
\begin{equation} 
-\left(\Psi_j + \Psi_{j + 2}\right) + b\Psi_{j + 1} = E\Psi_{j + 1}
\end{equation}
{The key to being able to solve this particular Schrodinger equation hinges upon the ability to exploit the periodicity of this configuration and use Bloch's theorem. Using Bloch's theorem, we write}
\begin{equation}
\Psi_k \left( \vec{r} \right) = u \left( \vec{r} \right) \phi \left( \vec{r} \right) = u \left( \vec{r} \right) e^{i \vec{k} \cdot \vec{r}} \nonumber
\end{equation}
{For the lattice geometry the solutions take the form}
\begin{equation}
\Psi_j = u \left( j \right) e^{i k j}
\end{equation}
\\
{where $u$ has the periodicity of the lattice.}

{For $j$ odd, let $u = u_1$, and for $j$ even, let $u = u_2$. Then the pair of equations arising from the application of the Schrodinger equation becomes}
\begin{equation}
-\left( u_2 e^{i k \left( j - 1 \right)} + u_2 e^{i k \left( j + 1 \right)} \right) + a u_1 e^{i k j } = E u_1 e^{i k j } \nonumber
\end{equation}
\begin{equation}
-\left( u_1 e^{i k j } + u_1 e^{i k \left( j + 2 \right)} \right) + b u_2 e^{i k \left(j + 1 \right) } = E u_1 e^{i k \left(j + 1 \right) } \nonumber
\end{equation}
{They can be simplified to become}
\begin{equation}
-2 u_2 cosk = \left( E - a \right) u_1 
\end{equation}
\begin{equation}
-2 u_1 cosk = \left( E - b \right) u_2 
\end{equation}
{Multiplying this pair of equations, Eqs (73) and (74), we obtain a single equation in terms of $E$, $a$, $b$ and $k$.}
\begin{equation}
\left( E - a \right)\left( E - b \right) = 4 cos^2k 
\end{equation}
{We can apply normalization $|u_1|^2 + |u_2|^2 = 1$ combined with Eqs (73) and (74) to arrive at expressions for $u_1$ and $u_2$.}
\begin{equation}
u_1 = \frac{\left(b - E\right)}{2cosk}\frac{e^{i\theta}}{\sqrt{1 + \frac{4cos^2k}{\left(E - a \right)^2}}}
\end{equation}
\begin{equation}
u_2 = \frac{e^{i\theta}}{\sqrt{1 + \frac{4cos^2k}{\left(E - a \right)^2}}} 
\end{equation}
{where $\theta$ is the phase.}

{We need to apply periodicity to express the form of $k$. Physically, periodicity arises because the quantum particle is isomorphic to a ring polymer, and so it must close upon itself. The periodicity is realized in the top and bottom rows of the original cyclic matrix formula by the inclusion of the -1's in the upper right and lower left corners:}
\\
\\
{Top Row:}
\begin{equation}
au_1e^{ik} - u_2e^{2ik} - e^{ikN}u_2 = Eu_1e^{ik}
\end{equation}
{Bottom Row:}
\begin{equation}
-u_1e^{ik} - u_1e^{ik\left(N - 1\right)} + bu_2e^{ikN} = Eu_2e^{ikN}
\end{equation}
{This leads to}
\begin{equation}
\left(E - a \right)u_1e^{ik} = -\left(e^{2ik} + e^{ikN}\right)u_2 
\end{equation}
\begin{equation}
-\left(e^{ik} + e^{ik\left(N - 1 \right)}\right)u_1 = \left(E - b\right)e^{ikN}u_2 
\end{equation}
{Let's manipulate the first equation out of the last pair of equations, Eq. (80). First, let's multiply it through by $e^{-ik}$.}
\begin{equation}
\left(E - a \right)u_1 = -\left(e^{ik} + e^{-ik}e^{ikN}\right)u_2 \nonumber
\end{equation}
{Comparing this equation with the very first equation derived for $u_1$, Eq. (73), we see that the following must be true for consistency:}
\begin{equation}
e^{ikN} = 1 \Rightarrow kN = 2\pi\nu \nonumber 
\end{equation}
{where $\nu$ is an integer. Thus, we get}
\begin{equation}
k = \frac{2 \pi \nu}{N}
\end{equation}
{Eq. (75) derived above is a quadratic in $E$, and it can be solved for $E$ to obtain}
\begin{equation}
E = \frac{a + b \pm \sqrt{\left(a - b \right)^2 + 16cos^2\left(\frac{2 \pi \nu}{N}\right)}}{2}
\end{equation}
{For simplicity, let's define the following variable:}
\begin{equation}
F_5\left(x\right) = \sqrt{\left(a - b \right)^2 + 16cos^2\left(x\right)} \nonumber
\end{equation}
{This radical is almost ubiquitous in what follows in this paper. For the case of discrete wavenumber, $x = \frac{2 \pi \nu}{N}$. In the limit $N \rightarrow \infty$, $x = u$.}

{Eq. (83) shows that the energies for the striped case configuration occur in two bands, one for the case of the radical being prepended by the plus sign, and one for the case of the minus sign. These two cases shall be referred to in this paper by $E_{\nu^+}$ and $E_{\nu^-}$, respectively.}
\subsection*{\textit{2. Partition function for the striped case}}
{Considering the two branches of energy dictated by Eq. (83), one can attain the following equation for the partition function for the striped case configuration, after a little algebra:}
\begin{equation}
Z = e^{-\beta\left(\frac{a + b}{2}\right)}  \sum_{\nu = 1}^N \left[e^{\frac{\beta}{2}F_5\left(\frac{2 \pi \nu}{N}\right)} + e^{-\frac{\beta}{2}F_5\left(\frac{2 \pi \nu}{N}\right)}\right] \nonumber
\end{equation} 

{As a confirmation, it can be shown that if we set $a = b = 2$, we obtain the free particle result.}

{Let's now consider $Z/N$.}
\begin{equation}
\frac{Z}{N} = \frac{e^{-\beta\left(\frac{a + b}{2}\right)}}{N} \sum_{\nu = 1}^N \left[e^{\frac{\beta}{2}F_5\left(\frac{2 \pi \nu}{N}\right)} + e^{-\frac{\beta}{2}F_5\left(\frac{2 \pi \nu}{N}\right)}\right] \nonumber
\end{equation}

{As usual, we perform a change of variable, then take the limit $N \rightarrow \infty$, which changes the summation to an integral, and we obtain}
\begin{equation}
\frac{Z}{N} = \frac{e^{-\beta\left(\frac{a + b}{2}\right)}}{\pi} \int_0^{2 \pi} cosh\left[\frac{\beta}{2}F_5\left( u \right)\right]du 
\end{equation}
{This integral, and all other integrals to follow in this paper, must be done numerically.}
\subsection*{\textit{3. Average energy for the striped case}}
{Let's calculate the average energy $\langle \hat{H} \rangle$}
\begin{equation}
\langle \hat{H} \rangle = \frac{\frac{1}{N}\langle \hat{H}e^{-\beta\hat{H}}\rangle}{\frac{Z}{N}} \nonumber
\end{equation}
{Taking into account both branches, the equation for the average energy takes the form}
\begin{equation}
\langle \hat{H} \rangle =  \nonumber
\end{equation}
\begin{equation}
\frac{e^{-\beta \left(\frac{a + b}{2}\right)}}{\frac{Z}{N}N}\sum_{\nu = 1}^N \left[\left( \frac{a + b}{2}\right) + \frac{1}{2}F_5\left( \frac{2 \pi \nu}{N} \right) \right]e^{-\frac{\beta}{2}F_5\left( \frac{2 \pi \nu}{N} \right)}  \nonumber
\end{equation}
\begin{equation}
 + \frac{e^{-\beta \left(\frac{a + b}{2}\right)}}{\frac{Z}{N}N}\sum_{\nu = 1}^N \left[\left( \frac{a + b}{2}\right) - \frac{1}{2}F_5\left( \frac{2 \pi \nu}{N} \right) \right] e^{\frac{\beta}{2}F_5\left( \frac{2 \pi \nu}{N} \right)} \nonumber
\end{equation}
{This equation can be simplified to become}
\begin{equation}
\langle \hat{H} \rangle = \frac{e^{-\beta \left(\frac{a + b}{2}\right)}}{\frac{Z}{N}N}\sum_{\nu = 1}^N\left[\left(a + b\right) cosh\left[ \frac{\beta}{2}F_5\left( \frac{2 \pi \nu}{N} \right) \right] \right] \nonumber
\end{equation}
\begin{equation}
-  \frac{e^{-\beta \left(\frac{a + b}{2}\right)}}{\frac{Z}{N}N}\sum_{\nu = 1}^N\left[ F_5\left( \frac{2 \pi \nu}{N} \right) sinh\left[ \frac{\beta}{2}F_5\left( \frac{2 \pi \nu}{N} \right) \right] \right] \nonumber
\end{equation}
{We perform a change of variable, then take the limit $N \rightarrow \infty$, which changes the summation to an integral, and we obtain}
\begin{equation}
\langle \hat{H} \rangle = \frac{a + b}{2}  - \frac{ \frac{1}{2}\int_0^{2 \pi} F_5 \left( u \right) sinh\left[ \frac{\beta}{2}F_5\left( u \right) \right] du}{\int_0^{2 \pi}  cosh\left[ \frac{\beta}{2}F_5 \left( u \right) \right]du }
\end{equation}
\subsection*{\textit{4. Average potential energy for the striped case}}
{The probability that the striped case system is in a state $\nu$ is}
\begin{equation}
P_{\nu} = \frac{e^{-\beta E_{\nu}}}{Z}
\end{equation}
{The average potential energy for state $\nu$ is given by}
\begin{equation}
\langle V \rangle_{\nu} = \frac{\sum_j \Psi_j^* V \Psi_j}{\sum_j|\Psi_j|^2}
\end{equation}
{For this problem, the distribution of the on-site potential energy from the atoms on the lattice is given by} 
\begin{equation}
\label{eq:mdiv}
V =
\begin{cases}
\epsilon & \text{j odd} \\
0 & \text{j even}
\end{cases}
\end{equation}
{and the wavefunctions are}
\begin{equation}
\label{eq:mdiv}
\Psi_j = e^{i k j}
\begin{cases}
u_1 & \text{j odd} \\
u_2 & \text{j even}
\end{cases}
\end{equation}
{This gives a $\nu$-dependent average potential energy of}
\begin{equation}
\langle V \rangle_{\nu} = \frac{|u_1|^2 \epsilon \frac{N}{2}}{\sum_j|\Psi_j|^2}
\end{equation}
{Because $\Psi_j = u_je^{i k j}$, we can write the complex square of $\Psi_j$ as}
\begin{equation}
|\Psi_j|^2 = |u_j|^2
\end{equation}
{We can then immediately obtain a normalization constant for $\Psi_j$}
\begin{equation}
\sum_j |\Psi_j|^2 = |u_1|^2 \frac{N}{2} + |u_2|^2 \frac{N}{2} \nonumber 
\end{equation}
\begin{equation}
= \left( |u_1|^2  + |u_2|^2 \right)\frac{N}{2} = \frac{N}{2}
\end{equation}
{Applying this normalization, we obtain the following formula for the $\nu$-dependent average potential energy}
\begin{equation}
\langle V \rangle_{\nu} = \epsilon |u_1|^2
\end{equation}
{Recall from Eq. (83) that there are two branches of E. We want to eventually calculate}
\begin{equation}
\langle V \rangle = \frac{\epsilon}{Z} \sum_{\nu}\left(|u_{1_{\nu^+}}|^2 e^{-\beta E_{\nu^+}} + |u_{1_{\nu^-}}|^2 e^{-\beta E_{\nu^-}}\right) \nonumber
\end{equation}
{Also, we can take the complex square of $u_1$, given by Eq. (76), and write $|u_{1_{\nu^+}}|^2$ and $|u_{1_{\nu^-}}|^2$, one for each branch of the frequency}
\begin{equation}
|u_{1_{\nu^+}}|^2 = \frac{\left(b - E_{\nu^+} \right)^2}{4cos^2\left(\frac{2 \pi \nu}{N}\right)} \left(\frac{\left(E_{\nu^+} - a \right)^2}{\left(E_{\nu^+} - a \right)^2 + 4cos^2\left(\frac{2 \pi \nu}{N}\right)}\right)\nonumber
\end{equation} 
{and}
\begin{equation}
|u_{1_{\nu^-}}|^2 = \frac{\left(b - E_{\nu^-} \right)^2}{4cos^2\left(\frac{2 \pi \nu}{N}\right)} \left(\frac{\left(E_{\nu^-} - a \right)^2}{\left(E_{\nu^-} - a \right)^2 + 4cos^2\left(\frac{2 \pi \nu}{N}\right)}\right) \nonumber
\end{equation}
{We need to work out expressions for $|u_{1_{\nu^+}}|^2$ and $|u_{1_{\nu^-}}|^2$ in terms of $\epsilon$ and $\nu$.}  
\\
\\
{In this problem, we consider the occupied sites to have on-site potential $\epsilon$ and the unoccupied sites to have 0 potential. To accomplish this, let $a = 2 + \epsilon$ and $b = 2$. After extensive algebraic calculations, we obtain}
\begin{equation}
|u_{1_{\nu^+}}|^2 = \frac{8cos^2\left(\frac{2 \pi \nu}{N}\right)}{16cos^2\left(\frac{2 \pi \nu}{N}\right) - \epsilon F_6 \left( \frac{2 \pi \nu}{N} \right) + \epsilon^2}
\end{equation}
{and}
\begin{equation}
|u_{1_{\nu^-}}|^2 = \frac{8cos^2\left(\frac{2 \pi \nu}{N}\right)}{16cos^2\left(\frac{2 \pi \nu}{N}\right) + \epsilon F_6 \left( \frac{2 \pi \nu}{N} \right) + \epsilon^2}
\end{equation}
{where, in the same spirit as $F_5\left( x \right)$, we create a new variable $F_6 \left( x \right)$ as follows:}
\begin{equation}
F_6\left( x \right) = \sqrt{\epsilon^2 + 16cos^2\left(x\right)} \nonumber
\end{equation}
{For the case of discrete wavenumber, $x = \frac{2 \pi \nu}{N}$. In the limit $N \rightarrow \infty$, $x = u$.}
\\
\\
{$\langle V \rangle$ now becomes}
\begin{equation}
\langle V \rangle = \frac{\epsilon}{Z} \sum_{\nu}\left[ \frac{8cos^2\left(\frac{2 \pi \nu}{N}\right)e^{-\frac{\beta}{2}\left(4 + \epsilon\right)}e^{-\frac{\beta}{2}F_6 \left( \frac{2 \pi \nu}{N} \right)}}{16cos^2\left(\frac{2 \pi \nu}{N}\right) - \epsilon F_6 \left( \frac{2 \pi \nu}{N} \right) + \epsilon^2}\right] \nonumber
\end{equation}
\begin{equation}
+ \frac{\epsilon}{Z} \sum_{\nu}\left[ \frac{8cos^2\left(\frac{2 \pi \nu}{N}\right)e^{-\frac{\beta}{2}\left(4 + \epsilon\right)}e^{\frac{\beta}{2}F_6 \left( \frac{2 \pi \nu}{N} \right)}}{16cos^2\left(\frac{2 \pi \nu}{N}\right) + \epsilon F_6 \left( \frac{2 \pi \nu}{N} \right) + \epsilon^2}\right] \nonumber
\end{equation}
{Making a change of variable, taking the limit as $N \rightarrow \infty$, and changing the summation to an integral, we get}
\begin{equation}
\langle V \rangle = \frac{ 4 \epsilon \int_0^{2 \pi} \frac{cos^2\left(u \right) e^{-\frac{\beta}{2}F_6\left( u \right)}}{16cos^2\left( u \right) - \epsilon F_6 \left( u \right) + \epsilon^2} du}{\int_0^{2 \pi} cosh\left[\frac{\beta}{2} F_6 \left( u \right)\right]  du} \nonumber
\end{equation} 
\begin{equation}
+ \frac{ 4 \epsilon \int_0^{2 \pi} \frac{cos^2\left(u \right) e^{\frac{\beta}{2}F_6\left( u \right)}}{16cos^2\left( u \right) + \epsilon F_6 \left( u \right) + \epsilon^2} du}{\int_0^{2 \pi} cosh\left[\frac{\beta}{2} F_6 \left( u \right)\right]  du} 
\end{equation}
\subsection*{\textit{5. Ground state energy for the striped case}}
{We are now in a position to predict analytically the ground state energy for the striped case configuration. The set of $E_{\nu^+}$ and $E_{\nu^-}$ given by Eq. (83) are all possible energy eigenvalues for this problem. For any $\nu$, $E_{\nu^-} < E_{\nu^+}$. The ground state energy occurs when $E_{\nu^-}$ is a minimum. This occurs when the radical is maximized, which occurs when $cos^2\left(\frac{2 \pi \nu}{N}\right)$ is maximized, which occurs for $\nu = \frac{N}{2}$. This leads to}
\begin{equation}
E_{ground} = \frac{a + b}{2} - \frac{1}{2}\sqrt{\left(a - b\right)^2 + 16} \nonumber
\end{equation}
{Let $\epsilon = 10.0$, $b = 2.0$, $a = b + \epsilon = 2.0 + 10.0 = 12.0$. From this, we get a value for the ground-state energy.}
\begin{equation}
E_{ground} \approx 1.6148
\end{equation}

{We can now theoretically predict the ground state potential energy for the striped configuration. We simply begin with Eq. (93) and substitute the form of $|u_1|^2$ from Eq. (76). Doing so, we can write}
\begin{equation}
\langle V \rangle_{\nu} = \epsilon \frac{\left(b - E\right)^2}{4cos^{2}k}\frac{1}{\left(1 + \frac{4cos^2k}{\left(E - a \right)^2}\right)} \nonumber
\end{equation}
{Using the above parameter settings and the value for $E_{ground}$ calculated in Eq. (97), we get}
\begin{equation}
\langle V \rangle_{ground} = (10.0) \frac{\left(2.0 - E_{ground}\right)^2}{4cos^{2}\pi} \nonumber 
\end{equation}
\begin{equation}
\times \frac{1}{\left(1 + \frac{4cos^{2}\pi}{\left(E_{ground} - 12.0 \right)^2}\right)} = 0.3577
\end{equation}
{We will later compare the computational results for $\langle V \rangle$ with Eq. (98).}
\subsection*{\textit{ 6. The density matrix for the striped case}}
{In Quantum Statistical Mechanics, one is typically interested in calculating the density matrix. $\left[15\right]$ The density matrix is a matrix that describes a quantum system in a mixed state, as opposed to one in a pure state, which would simply be described by a single state vector. The density matrix is a quantum-mechanical analogue to the phase-space probability in Classical Statistical Mechanics. Explicitly, suppose that a given state $\ket{\psi}$ may be found in state $\ket{\psi_1}$ with probability $p_1$, in state $\ket{\psi_2}$ with probability $p_2$, $\ldots$, in state $\ket{\psi_n}$ with probability $p_n$. The density operator for this system is then}
\begin{equation}
\hat{\rho} = \sum_i p_i \ket{\psi_i}\bra {\psi_i}
\end{equation}
{By choosing a basis $\ket{u_m}$, which does not even need to be orthogonal, one may resolve the density operator into a density matrix, which has the matrix elements}
\begin{equation}
\rho_{mn} = \sum_i p_i \braket{u_m}{\psi_i} \braket{\psi_i}{u_n}
\end{equation}  
{Then, for a given operator $\hat{A}$, the expectation value $\langle A \rangle$ is given by}
\begin{equation}
\langle A \rangle = \sum_i p_i \bra{\psi_i} \hat{A} \ket{\psi_i} = \sum_n \bra{u_n} \hat{\rho} \hat{A} \ket{u_n} = Tr\left(\hat{\rho} \hat{A} \right)
\end{equation}
{The expectation value of A for the mixed state is the sum of the expectation values of for each of the pure states $\ket{\psi_i}$ weighted by probabilities $p_i$.}

{For any atomic configuration, the density matrix is an important mathematical object to calculate. In the case of a configuration of atoms on a one-dimensional lattice, the density matrix can be employed to provide a correlation function for the quantum particle. Let's begin computing the density matrix for the striped case configuration.}
\\
{First, let's write}
\begin{equation}
\rho = \langle \phi^{*} \left( \vec{x} \right) \phi \left( \vec{x}^{'} \right) \rangle = \frac{\sum_{\nu} \phi_{\nu}^{*} \left( \vec{x} \right) \phi_{\nu} \left( \vec{x}^{'} \right) e^{-\beta E_{\nu}}}{Z} \nonumber
\end{equation}
{Expressing this relation in the formalism of the lattice, we write}
\begin{equation}
\langle \phi^* \left( j \right) \phi \left( j' \right) \rangle = \frac{\sum_{\nu \text{(states)}} \phi_{\nu, j}^* \phi_{\nu, j'} e^{-\beta E_{\nu}}}{Z} \nonumber
\end{equation}
{For each $k$, we have $E_+ \left( k \right)$ and $E_- \left( k \right)$.} 
\\
{Applying normalization, we write}
\begin{equation}
\frac{\left(\frac{2}{N}\right) \sum_{\text{states}} u_j^* \left( \nu \right) u_{j'} \left( \nu \right) e^{-\beta E_{\nu}} e^{i k \left( j' - j \right)}}{Z} = D_{jj'} \nonumber
\end{equation}
{which is matrix element $j$, $j'$ of density matrix $D$.}
\begin{equation} 
D_{jj'} = \frac{\left(\frac{2}{N}\right) \sum_{\nu^+} u_j^*\left(\nu^+\right)  u_{j'}\left(\nu^+\right)e^{-\beta E_{\nu^+}}e^{i k  \left( j' - j \right)}}{Z} \nonumber
\end{equation}
\begin{equation}
+ \frac{\left(\frac{2}{N}\right) \sum_{\nu^-} u_j^*\left(\nu^-\right)  u_{j'}\left(\nu^-\right)e^{-\beta E_{\nu^-}}e^{i k  \left( j' - j \right)}}{Z}
\end{equation}
\\
{The $D_{jj'}$ matrix can be separated into a $D_{jj'}^+$ and a $D_{jj'}^-$, given by}
\begin{equation}
D_{jj'}^+ = \frac{2}{N}\frac{1}{Z} \sum_{\nu^+} u_j^*\left( \nu^+ \right) u_{j'}\left( \nu^+ \right)e^{-\beta E_{\nu^+}}e^{i k \left( j' - j \right)}
\end{equation}
\begin{equation}
D_{jj'}^- = \frac{2}{N}\frac{1}{Z} \sum_{\nu^-} u_j^*\left( \nu^- \right) u_{j'}\left( \nu^- \right)e^{-\beta E_{\nu^-}}e^{i k \left( j' - j \right)}
\end{equation}
{Let's look at the N $\times$ N matrix $D_{jj'}^+$ first. It is displayed in its expanded form below, multiplied by $\frac{N}{2} Z$.}
\begin{widetext}
\begin{equation}
\textbf{$\frac{N}{2} Z D_{jj'}^+$} = 
\begin{pmatrix}
\sum_{\nu^+}u_1^*u_1e^{-\beta E_{\nu^+}}, \hfill e^{-ik}\sum_{\nu^+}u_1^*u_2e^{-\beta E_{\nu^+}}, \hfill \cdots \hfill e^{(2 - N)ik}\sum_{\nu^+}u_1^*u_1e^{-\beta E_{\nu^+}}, \hfill e^{(1 - N)ik}\sum_{\nu^+}u_1^*u_2e^{-\beta E_{\nu^+}} \\
e^{ik}\sum_{\nu^+}u_2^*u_1e^{-\beta E_{\nu^+}}, \hfill \sum_{\nu^+}u_2^*u_2e^{-\beta E_{\nu^+}}, \hfill \cdots \hfill e^{(3-N)ik}\sum_{\nu^+}u_2^*u_1e^{-\beta E_{\nu^+}}, \hfill e^{(2-N)ik}\sum_{\nu^+}u_2^*u_2e^{-\beta E_{\nu^+}} \\
\vdots \\
\vdots \\
e^{(N-2)ik}\sum_{\nu^+}u_1^*u_1e^{-\beta E_{\nu^+}}, \hfill e^{(N-3)ik}\sum_{\nu^+}u_1^*u_2e^{-\beta E_{\nu^+}}, \hfill \cdots \hfill \sum_{\nu^+}u_1^*u_1e^{-\beta E_{\nu^+}}, \hfill e^{-ik}\sum_{\nu^+}u_1^*u_2e^{-\beta E_{\nu^+}} \\
e^{(N-1)ik}\sum_{\nu^+}u_2^*u_1e^{-\beta E_{\nu^+}}, \hfill e^{(N-2)ik}\sum_{\nu^+}u_2^*u_2e^{-\beta E_{\nu^+}}, \hfill \cdots \hfill e^{ik}\sum_{\nu^+}u_2^*u_1e^{-\beta E_{\nu^+}}, \hfill \sum_{\nu^+}u_2^*u_2e^{-\beta E_{\nu^+}} \\
\end{pmatrix} \nonumber 
\end{equation} 
\end{widetext}
{This matrix is constructed understanding that $u_j = u_1$ if $j$ is odd and that $u_j = u_2$ if $j$ is even.}
\\
\\
{$D_{jj'}^-$ is the same matrix as $D_{jj'}^+$ except that everywhere one sees $\nu^+$ in $D_{jj'}^+$, one must substitute it for $\nu^-$ in $D_{jj'}^-$.}
\\
\\
{In the expanded display of the matrix $D_{jj'}^+$, $u_1 = u_{1_{\nu^+}}$ and $u_2 = u_{2_{\nu^+}}$. For the matrix $D_{jj'}^-$, $u_1 = u_{1_{\nu^-}}$ and $u_2 = u_{2_{\nu^-}}$.}
\\
\\
{We have expressions for $|u_{1_{\nu^+}}|^2$ and $|u_{1_{\nu^-}}|^2$, Eqs (94) and (95), respectively. We can easily derive expressions for $|u_{2_{\nu^+}}|^2$ and $|u_{2_{\nu^-}}|^2$. Utilizing the normalization condition, we can write}
\begin{equation}
|u_{2_{\nu^+}}|^2 = 1 - |u_{1_{\nu^+}}|^2  \nonumber
\end{equation}
{and substituting Eq. (94) we obtain} 
\begin{equation}
|u_{2_{\nu^+}}|^2 = 1 - \frac{8cos^2\left(\frac{2 \pi \nu}{N}\right)}{16cos^2\left(\frac{2 \pi \nu}{N}\right) - \epsilon F_6 \left( \frac{2 \pi \nu}{N} \right) + \epsilon^2} \nonumber
\end{equation}
{This simplifies to a final form for $|u_{2_{\nu^+}}|^2$} 
\begin{equation}
|u_{2_{\nu^+}}|^2 = \frac{8cos^2\left(\frac{2 \pi \nu}{N}\right) - \epsilon F_6 \left( \frac{2 \pi \nu}{N} \right) + \epsilon^2 }{16cos^2\left(\frac{2 \pi \nu}{N}\right) - \epsilon F_6 \left( \frac{2 \pi \nu}{N} \right) + \epsilon^2} 
\end{equation}
{Likewise, for $|u_{2_{\nu^-}}|^2$, we begin with the normalization condition}
\begin{equation}
|u_{2_{\nu^-}}|^2 = 1 - |u_{1_{\nu^-}}|^2  \nonumber
\end{equation}
{and end up with}
\begin{equation}
|u_{2_{\nu^-}}|^2 = \frac{8cos^2\left(\frac{2 \pi \nu}{N}\right) + \epsilon F_6 \left( \frac{2 \pi \nu}{N} \right) + \epsilon^2 }{16cos^2\left(\frac{2 \pi \nu}{N}\right) + \epsilon F_6 \left( \frac{2 \pi \nu}{N} \right) + \epsilon^2}
\end{equation}
{Returning to the $D_{jj'}$ matrix, all of the multiplications of $u_j^* u_{j'}$ come down to}
\begin{equation}
u_{1_{\nu^+}}^* \cdot u_{1_{\nu^+}} = |u_{1_{\nu^+}}|^2  \nonumber
\end{equation}
\begin{equation}
u_{2_{\nu^+}}^* \cdot u_{2_{\nu^+}} = |u_{2_{\nu^+}}|^2  \nonumber
\end{equation}
{which we have already calculated, and}
\begin{equation}
u_{1_{\nu^+}}^* \cdot u_{2_{\nu^+}} = u_{2_{\nu^+}}^* \cdot u_{1_{\nu^+}} \nonumber
\end{equation}
\\
\begin{equation}
u_{1_{\nu^+}}^* \cdot u_{2_{\nu^+}} =  \nonumber 
\end{equation}
\begin{equation}
\frac{2\sqrt{2}cos\left( \frac{2 \pi \nu}{N} \right)\left[ 8cos^2\left(\frac{2 \pi \nu}{N}\right) - \epsilon F_6 \left( \frac{2 \pi \nu}{N} \right) + \epsilon^2 \right]^\frac{1}{2}}{16cos^2\left(\frac{2 \pi \nu}{N}\right) - \epsilon F_6 \left( \frac{2 \pi \nu}{N} \right) + \epsilon^2}
\end{equation}
{Likewise,}
\begin{equation}
u_{1_{\nu^-}}^* \cdot u_{1_{\nu^-}} = |u_{1_{\nu^-}}|^2  \nonumber
\end{equation}
\begin{equation}
u_{2_{\nu^-}}^* \cdot u_{2_{\nu^-}} = |u_{2_{\nu^-}}|^2  \nonumber
\end{equation}
{which we have already calculated, and}
\begin{equation}
u_{1_{\nu^-}}^* \cdot u_{2_{\nu^-}} = u_{2_{\nu^-}}^* \cdot u_{1_{\nu^-}} \nonumber
\end{equation}
\begin{equation}
u_{1_{\nu^-}}^* \cdot u_{2_{\nu^-}} =  \nonumber 
\end{equation}
\begin{equation}
\frac{2\sqrt{2}cos\left( \frac{2 \pi \nu}{N} \right)\left[ 8cos^2\left(\frac{2 \pi \nu}{N}\right) + \epsilon F_6 \left( \frac{2 \pi \nu}{N} \right) + \epsilon^2 \right]^\frac{1}{2}}{16cos^2\left(\frac{2 \pi \nu}{N}\right) + \epsilon F_6 \left( \frac{2 \pi \nu}{N} \right) + \epsilon^2}
\end{equation}
{We want to next convert the sums to integrals. Going through the same process as with the potential energy $\langle V \rangle$, one ends up getting an integral from 0 to $2 \pi$ in taking the limit $N \rightarrow \infty$, changing the variable to $u$, and multiplying the new integral by $\frac{N}{2 \pi}$.}
\\
In doing so, we need to calculate three different integrals for $D^+$ and three different integrals for $D^-$.}
\\
\\
{\bf{First $D^+$ Integral:}}
\begin{equation}
\frac{1}{\pi \left( \frac{Z}{N}\right)} \int_0^{2 \pi} |u_{1_{\nu^+}}\left( u \right)|^2 e^{-\beta E_{\nu^+}\left( u \right)}e^{i u \left( j - j'\right)} du  \nonumber 
\end{equation}
\begin{equation}
=  \frac{8}{\pi} e^{-\beta\left( \frac{a + b}{2}\right)} \left( \frac{Z}{N}\right)^{-1} \nonumber
\end{equation}
\begin{equation}
\times \int_0^{2 \pi} \frac{cos \left[u \left( j - j' \right)\right] cos^2 u e^{-\frac{\beta}{2}F_6 \left( u \right)}}{16cos^2 u - \epsilon F_6 \left( u \right) + \epsilon^2} du
\end{equation}
\\
{where $\left( \frac{Z}{N} \right)$ is given by Eq. (84).}
\\
\\
{\bf{Second $D^+$ Integral:}}
\begin{equation}
\frac{1}{\pi \left( \frac{Z}{N}\right)} \int_0^{2 \pi} |u_{2_{\nu^+}}\left( u \right)|^2 e^{-\beta E_{\nu^+}\left( u \right)}e^{i u \left( j - j'\right)} du  \nonumber 
\end{equation}
\begin{equation}
 = \frac{1}{\pi} e^{-\beta\left( \frac{a + b}{2}\right)} \left( \frac{Z}{N}\right)^{-1} \nonumber
\end{equation}
\begin{equation}
\times \int_0^{2 \pi} \frac{cos \left[u \left( j - j' \right)\right] \left[ 8cos^2 u -\epsilon F_6\left( u \right) + \epsilon^2\right]}{16cos^2 u - \epsilon F_6 \left( u \right) + \epsilon^2} \nonumber
\end{equation}
\begin{equation}
\times e^{-\frac{\beta}{2} F_6 \left( u \right)} du
\end{equation}
\\
{\bf{Third $D^+$ Integral:}}
\begin{equation}
\frac{1}{\pi \left( \frac{Z}{N}\right)} \int_0^{2 \pi} u_{1_{\nu^+}}^*\left( u \right) u_{2_{\nu^+}}\left( u \right) e^{-\beta E_{\nu^+}\left( u \right)}e^{i u \left( j - j'\right)} du  \nonumber
\end{equation}
\begin{equation}
 = \frac{1}{\pi} e^{-\beta\left( \frac{a + b}{2}\right)} \left( \frac{Z}{N}\right)^{-1} \int_0^{2 \pi} \frac{cos \left[u \left( j - j' \right)\right] \left(2\sqrt{2}cosu \right)}{16cos^2 u - \epsilon F_6 \left( u \right) + \epsilon^2} \nonumber
\end{equation}
\begin{equation}
\times \left[ 8cos^2 u -\epsilon F_6 \left( u \right) + \epsilon^2\right]^\frac{1}{2} e^{-\frac{\beta}{2} F_6 \left( u \right)} du
\end{equation}
\\
{\bf{First $D^-$ Integral:}}
\begin{equation}
\frac{1}{\pi \left( \frac{Z}{N}\right)} \int_0^{2 \pi} |u_{1_{\nu^-}}\left( u \right)|^2 e^{-\beta E_{\nu^-}\left( u \right)}e^{i u \left( j - j'\right)} du \nonumber
\end{equation}
\begin{equation}
 = \frac{8}{\pi} e^{-\beta\left( \frac{a + b}{2}\right)} \left( \frac{Z}{N}\right)^{-1} \nonumber
\end{equation}
\begin{equation}
\times \int_0^{2 \pi} \frac{cos \left[u \left( j - j' \right)\right] cos^2 u e^{\frac{\beta}{2} F_6\left( u \right)}}{16cos^2 u + \epsilon F_6 \left( u \right) + \epsilon^2} du
\end{equation}
\\
{\bf{Second $D^-$ Integral:}}
\begin{equation}
\frac{1}{\pi \left( \frac{Z}{N}\right)} \int_0^{2 \pi} |u_{2_{\nu^-}}\left( u \right)|^2 e^{-\beta E_{\nu^-}\left( u \right)}e^{i u \left( j - j'\right)} du  \nonumber 
\end{equation}
\begin{equation}
 = \frac{1}{\pi} e^{-\beta\left( \frac{a + b}{2}\right)} \left( \frac{Z}{N}\right)^{-1} \nonumber
\end{equation}
\begin{equation}
\times \int_0^{2 \pi} \frac{cos \left[u \left( j - j' \right)\right] \left[ 8cos^2 u +\epsilon F_6\left( u \right) + \epsilon^2\right]}{16cos^2 u + \epsilon\ F_6 \left( u \right) + \epsilon^2} \nonumber
\end{equation}
\begin{equation}
\times e^{\frac{\beta}{2} F_6 \left( u \right)} du
\end{equation}
\\
{\bf{Third $D^-$ Integral:}}
\begin{equation}
\frac{1}{\pi \left( \frac{Z}{N}\right)} \int_0^{2 \pi} u_{1_{\nu^-}}^*\left( u \right) u_{2_{\nu^-}}\left( u \right) e^{-\beta E_{\nu^-}\left( u \right)} e^{i u \left( j - j'\right)} du  \nonumber
\end{equation}
\begin{equation}
 = \frac{1}{\pi} e^{-\beta\left( \frac{a + b}{2}\right)} \left( \frac{Z}{N}\right)^{-1}  \int_0^{2 \pi} \frac{cos \left[u \left( j - j' \right)\right] \left(2\sqrt{2}cosu \right)}{16cos^2 u + \epsilon F_6 \left( u \right) + \epsilon^2}\nonumber
\end{equation}
\begin{equation}
\times \left[ 8cos^2 u +\epsilon F_6 \left( u \right) + \epsilon^2\right]^\frac{1}{2} e^{\frac{\beta}{2} F_6 \left( u \right)} du
\end{equation}
\subsection*{B. Monte Carlo calculation for the striped case}
\subsection*{\textit{1. Metropolis sampling}}
{The probability of a specific walk on the lattice is proportional to}
\begin{equation}
exp\left(-\frac{\beta V}{p}\right)\left[\prod_{\alpha = 1}^p I_{s_{\alpha}}\left(\frac{2 \beta t}{p}\right)\right]\Delta\left(\sum_{\alpha = 1}^p s_{\alpha} \right) \nonumber
\end{equation}
{Thus all averages must now include the Gibbs factor $exp\left(-\frac{\beta V}{p}\right)$ as well as the product of modified Bessel functions. In contrast with the free particle, in the general interacting system the presence of this factor in the distribution function prevents us from being able to directly sample the probability distribution for the random walk. To deal with this complication, we employ what is called Metropolis sampling. That is, we deal with the free particle conditional probability, as expressed previously, to generate a walk, but we then employ rejection to produce a sequence of walks which satisfies the correct distribution. Let $q$ represent the acceptance factor, }
\begin{equation}
q = \frac{\prod_{\alpha = 1}^p e^{-\beta V'_{j_{\alpha}}}}{\prod_{\alpha = 1}^p e^{-\beta V_{j_{\alpha}}}} = \frac{exp\left(-\frac{\beta}{p}\sum_{\alpha = 1}^p V'_{j_{\alpha}}\right)}{exp\left(-\frac{\beta}{p}\sum_{\alpha = 1}^p V_{j_{\alpha}}\right)}
\end{equation}
{where $V'$ is the potential energy of the new walk and $V$ is the potential energy of the previous walk. Then, according to the Metropolis criteria, if $q > 1$ we automatically accept the new walk, while if $q < 1$ we only accept the new walk with probability $q$. This is determined by drawing a random number on the unit interval. If the random number is less than $q$, we reject the new walk and we hold onto the previous walk for the next iteration.}

{Earlier we discussed the generation of the random walks using a conditional probability argument. Originally, the PIMC computer program was designed to run for a specified number of iterations such that for each iteration a new sequence of pseudo-particle positions based on the conditional probability is proposed which will replace the current sequence of pseudo-particle positions if the Metropolis condition is satisfied. It turns out that if one considers total replacement of the closed chain for each iteration the acceptance percentage is too low. Essentially, this is an indicator that the statistics are not of sufficient quality to perform stochastic processes. We need to find a way to sufficiently slow the changes of the closed-chain qp from iteration to iteration to improve this acceptance percentage. We chose to consider replacing randomly selected segments of the chain of specified size from iteration to iteration. We tried segments of size equal to 20$\%$, 50$\%$ and 90$\%$ of the total chain. After performing several trials it was determined that we get the best acceptance percentages if we perform segment replacements of a size equal to 20$\%$ of the length of the chain. All of the PIMC results shown in this paper for both the free quantum particle and for the quantum particle interacting with the striped configuration of atoms are due to runs where a segment replacement size of 20$\%$ was imposed. } 
\subsection*{\textit{2. Form of the operators}}
\begin{figure} [h]
\begin{center}
\includegraphics[scale=0.7] {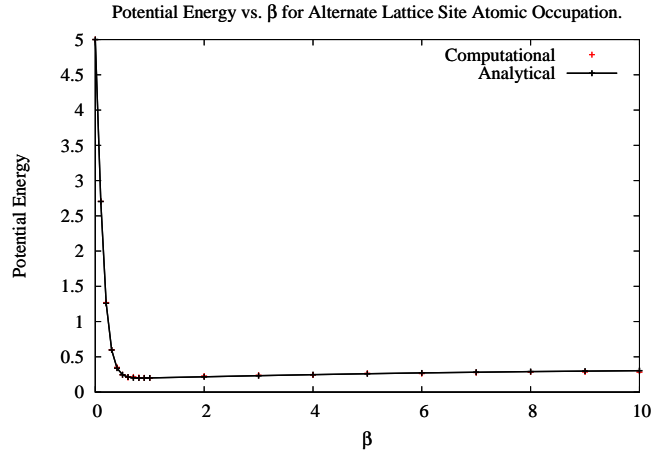} 
\caption{\label {fig:qm/complexfunctions}  Analytical and Computational Average Potential Energy versus inverse temperature, $\beta$, for a quantum particle moving on the lattice with a striped configuration. }
\end{center}
\end{figure}

\begin{figure} [h]
\begin{center}
\includegraphics[scale=0.65] {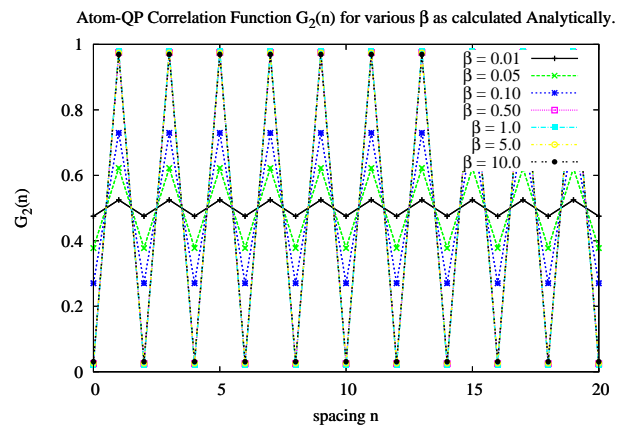} 
\caption{\label {fig:qm/complexfunctions}  The Atom-Quantum-Particle Correlation plots calculated analytically for various values of the inverse temperature $\beta$ for the Striped Configuration. Note: The drawn lines have no physical significance. Their purpose is to aid the eye in viewing the data trends. }
\end{center}
\end{figure}

\begin{figure} [h]
\begin{center}
\includegraphics[scale=0.65] {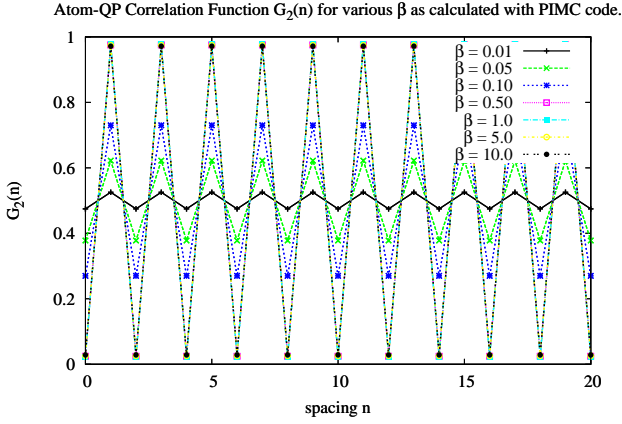} 
\caption{\label {fig:qm/complexfunctions}  The Atom-Quantum-Particle Correlation plots calculated through Monte Carlo simulations for various values of the inverse temperature $\beta$ for the Striped Configuration. Note: The drawn lines have no physical significance. Their purpose is to aid the eye in viewing the data trends. }
\end{center}
\end{figure}

\begin{figure} [h]
\begin{center}
{\includegraphics[scale=0.65] {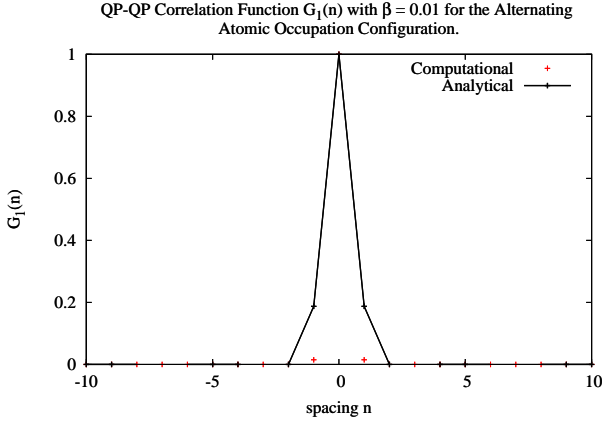} }
\caption{\label {fig:qm/complexfunctions} The Quantum Particle - Quantum Particle Correlation Function for the Striped Configuration for $\beta = 0.01$. Note: The drawn lines in this figure and in Figures 7 - 12 have no physical significance. Their purpose is to aid the eye in viewing the data trends.}
\end{center}
\end{figure}

\begin{figure} [h]
\begin{center}
\includegraphics[scale=0.65] {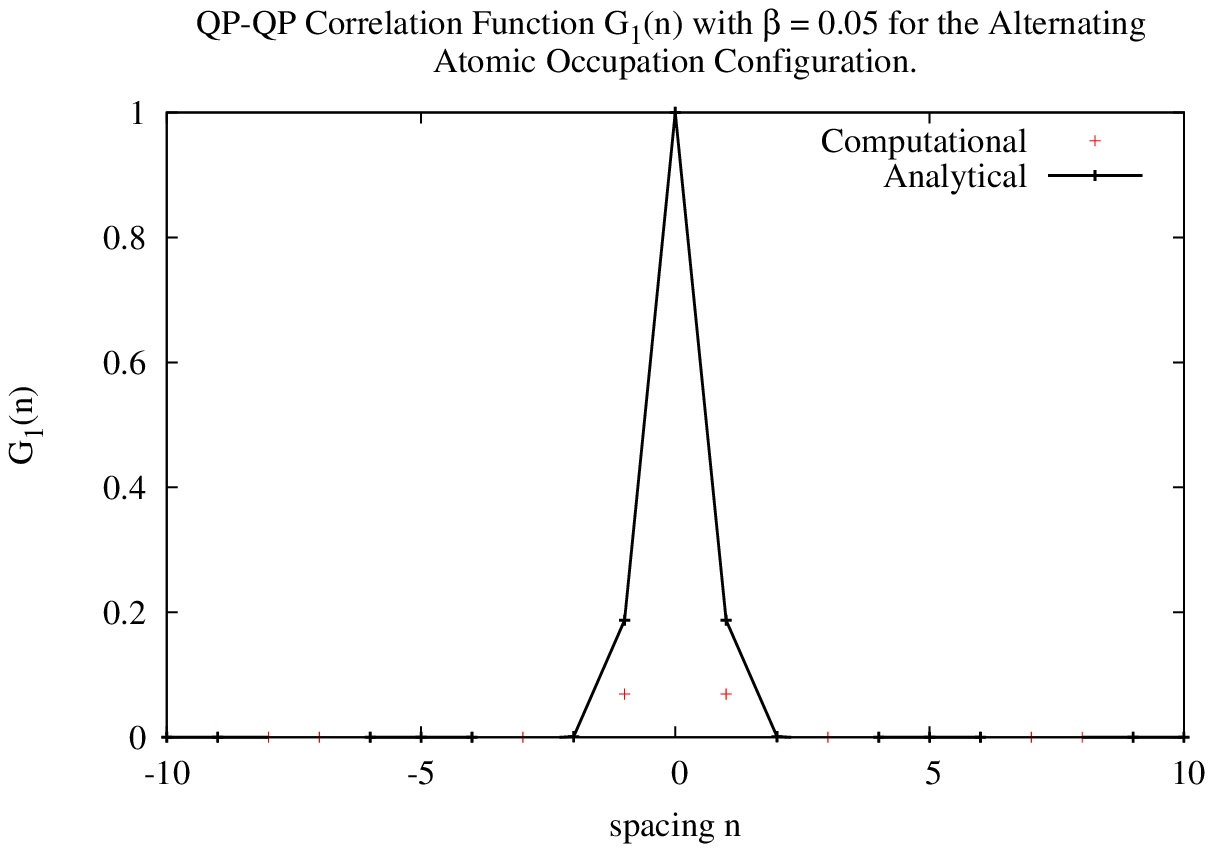} 
\caption{\label {fig:qm/complexfunctions}  The Quantum Particle - Quantum Particle Correlation Function for the Striped Configuration for $\beta = 0.05$. }
\end{center}
\end{figure}

\begin{figure} [h]
\begin{center}
\includegraphics[scale=0.65] {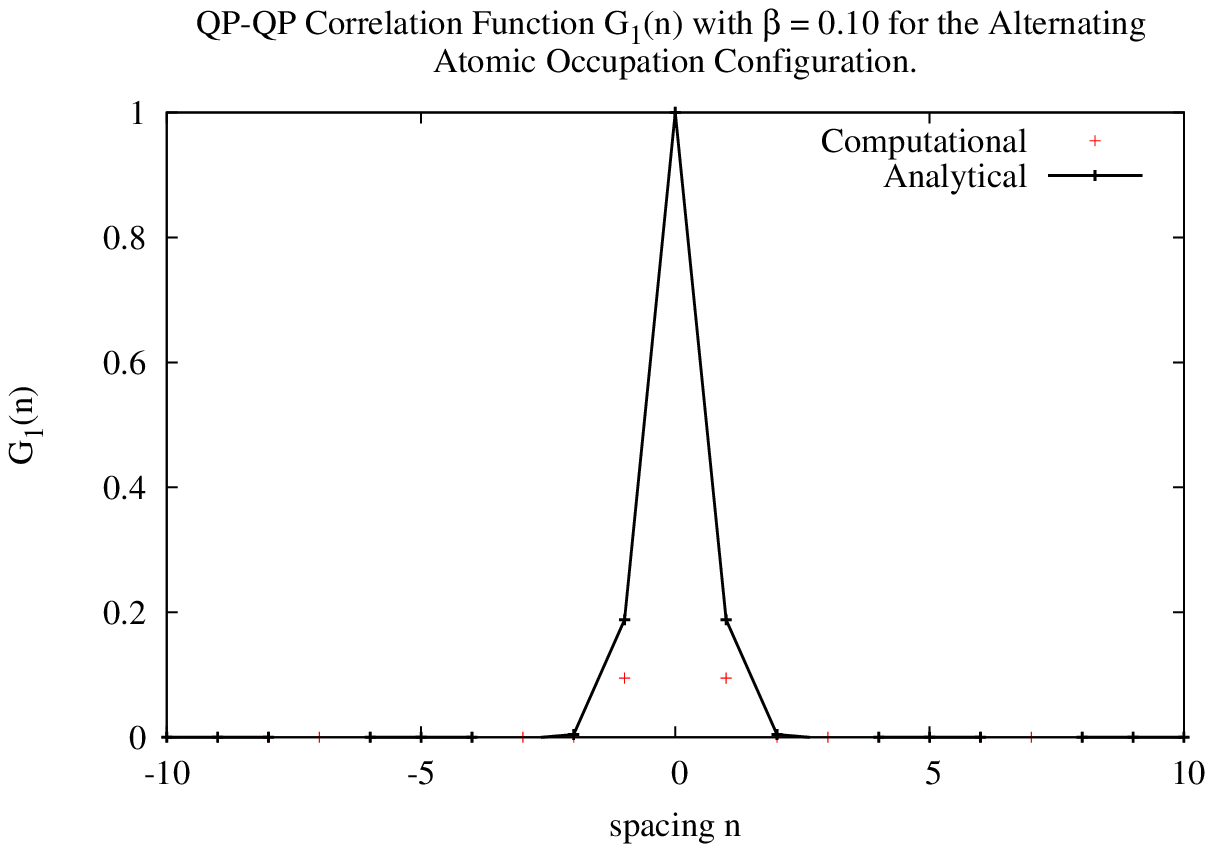} 
\caption{\label {fig:qm/complexfunctions} The Quantum Particle - Quantum Particle Correlation Function for the Striped Configuration for $\beta = 0.10$. }
\end{center}
\end{figure}

\begin{figure} [h]
\begin{center}
\includegraphics[scale=0.65] {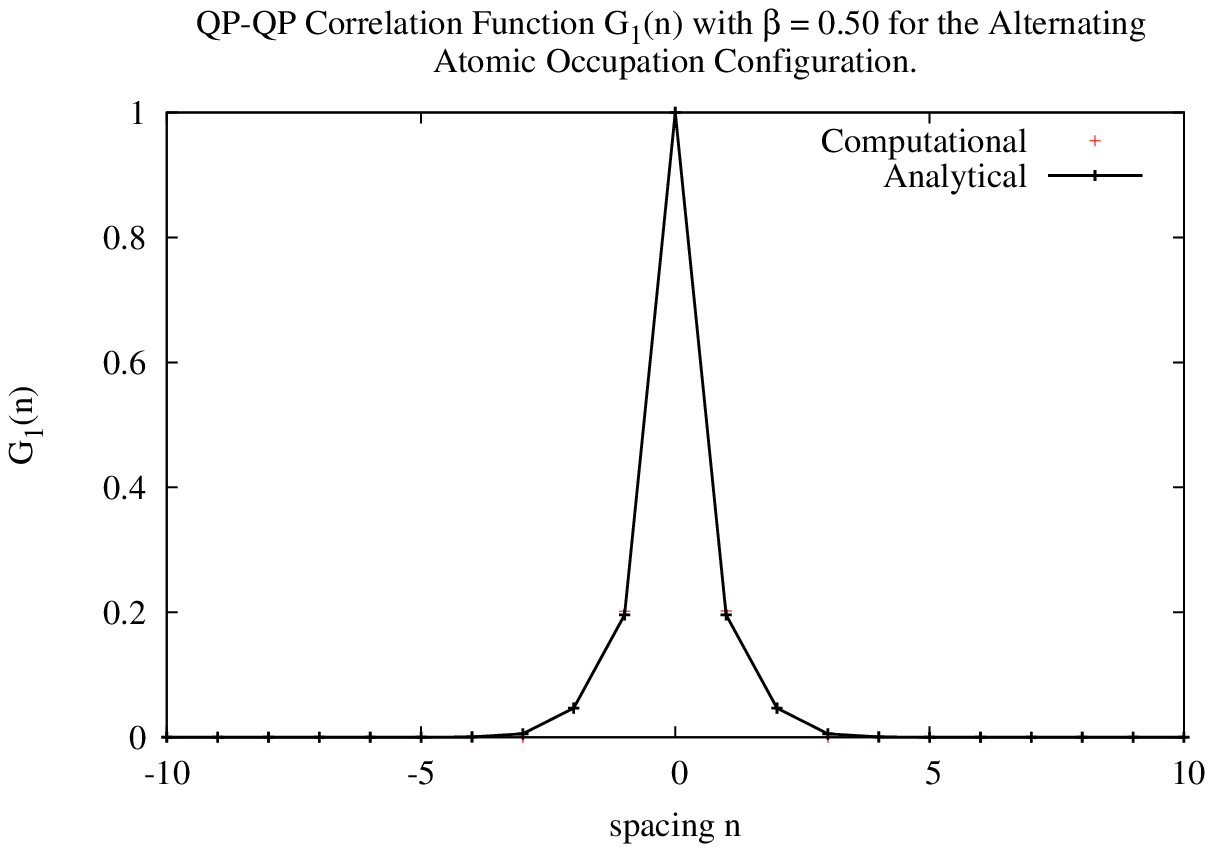} 
\caption{\label {fig:qm/complexfunctions}  The Quantum Particle - Quantum Particle Correlation Function for the Striped Configuration for $\beta = 0.50$. }
\end{center}
\end{figure}

\begin{figure} [h]
\begin{center}
\includegraphics[scale=0.65] {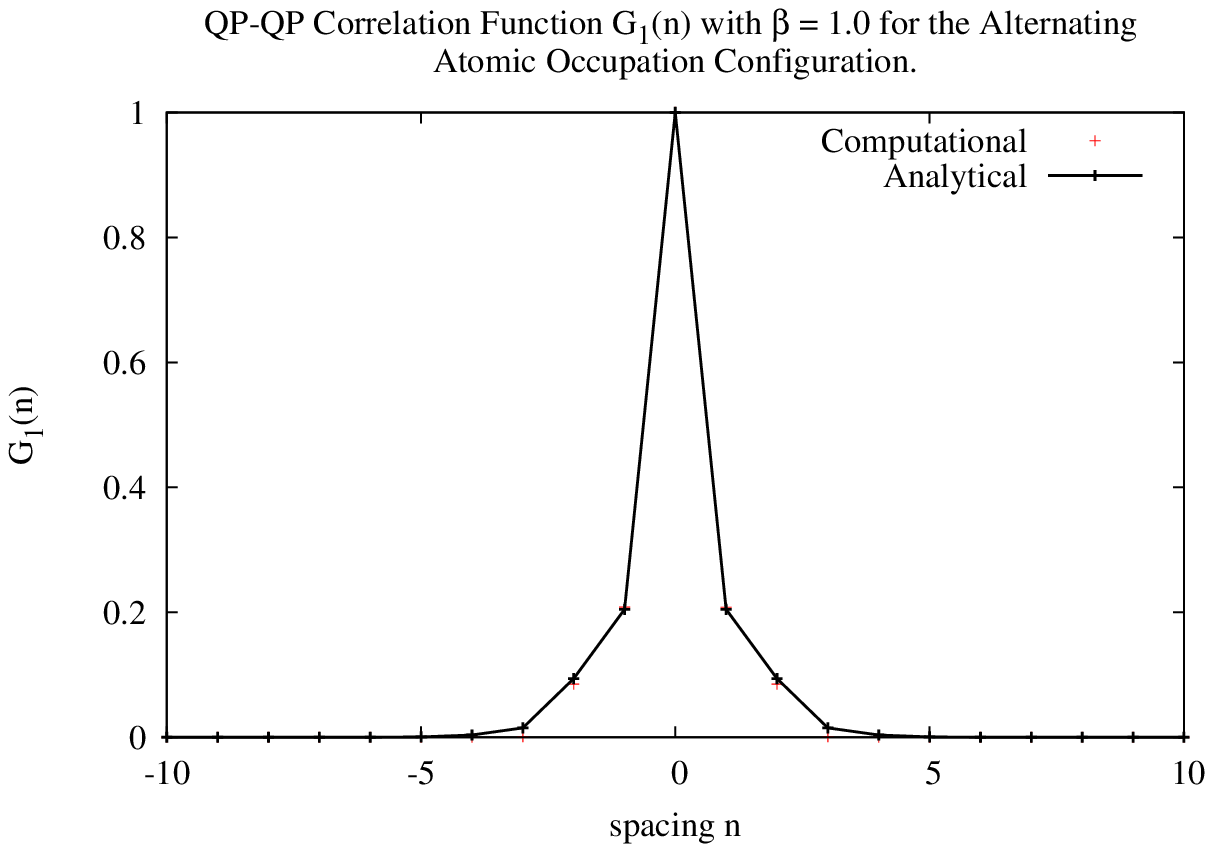} 
\caption{\label {fig:qm/complexfunctions} The Quantum Particle - Quantum Particle Correlation Function for the Striped Configuration for $\beta = 1.0$. }
\end{center}
\end{figure}

\begin{figure} [h]
\begin{center}
\includegraphics[scale=0.65] {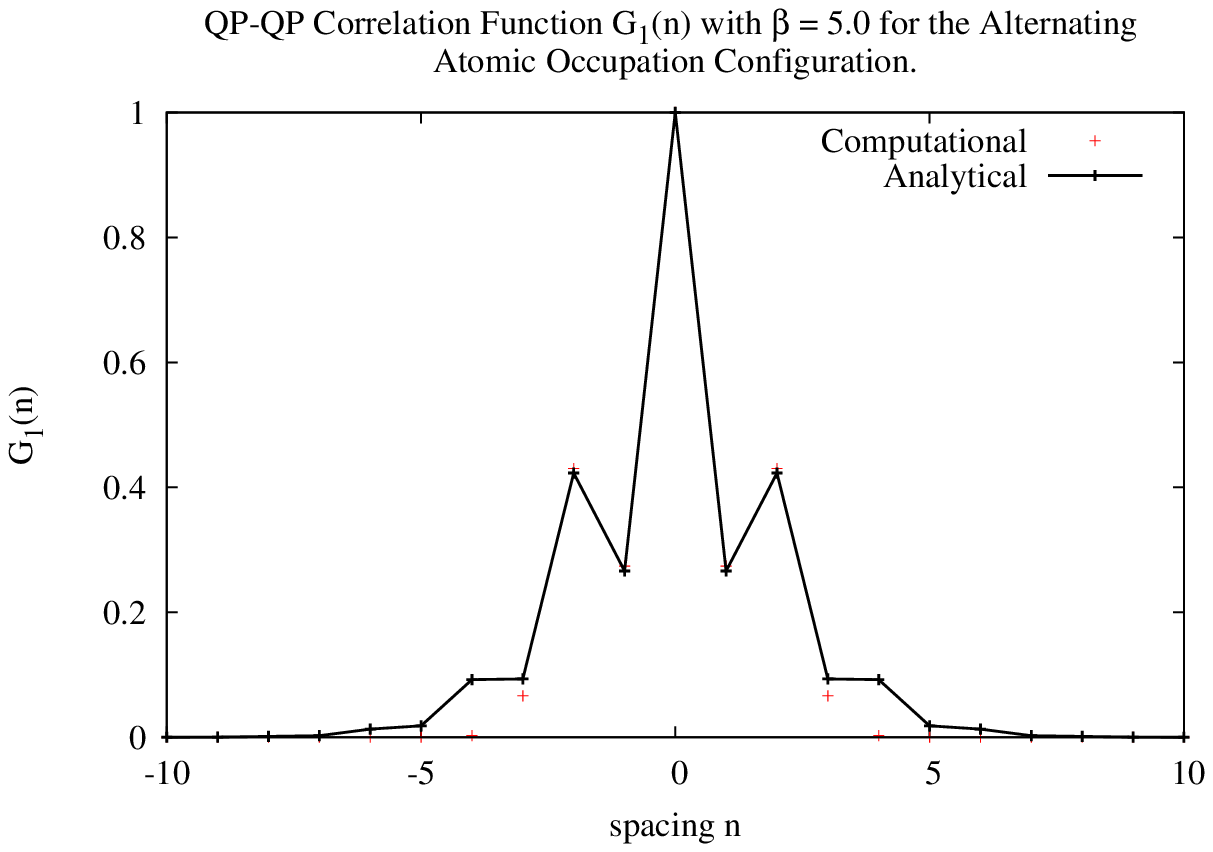} 
\caption{\label {fig:qm/complexfunctions}  The Quantum Particle - Quantum Particle Correlation Function for the Striped Configuration for $\beta = 5.0$. }
\end{center}
\end{figure}

\begin{figure} [h]
\begin{center}
\includegraphics[scale=0.65] {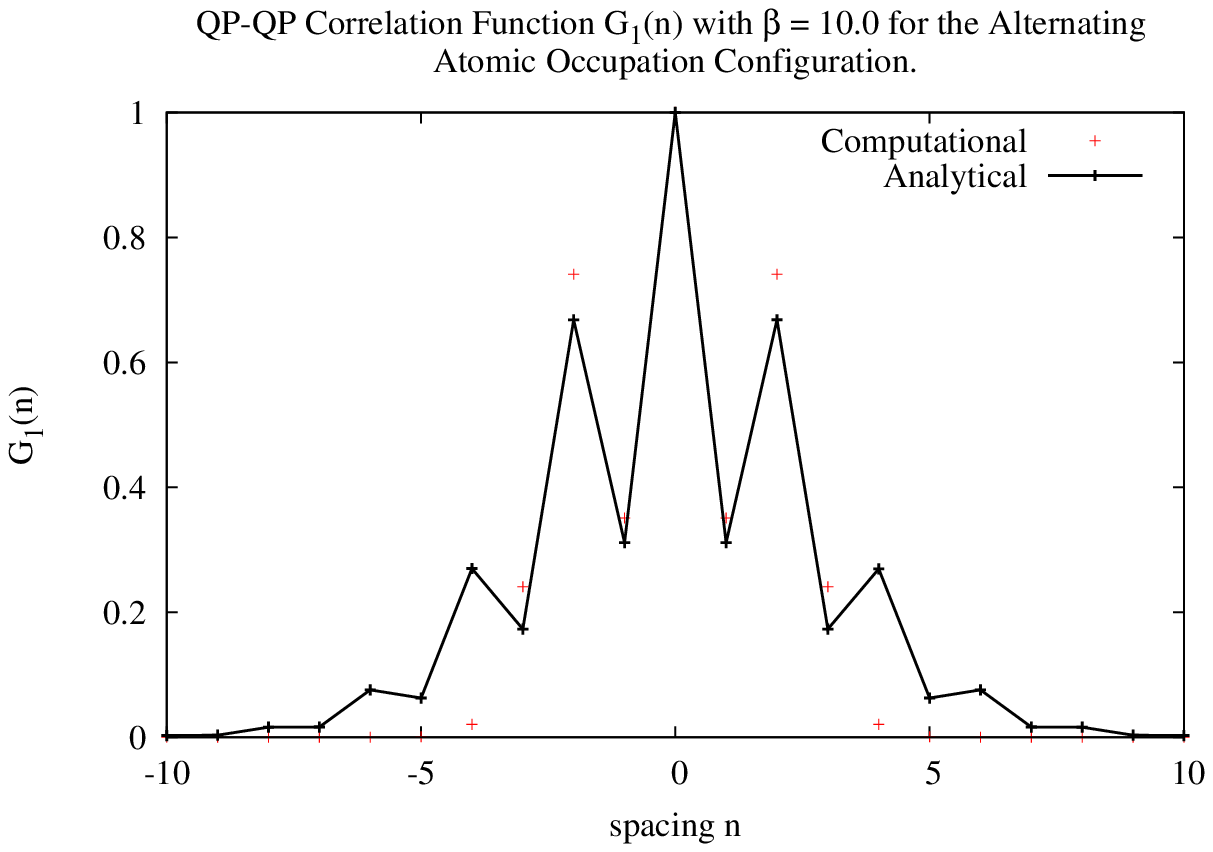} 
\caption{\label {fig:qm/complexfunctions}  The Quantum Particle - Quantum Particle Correlation Function for the Striped Configuration for $\beta = 10.0$. }
\end{center}
\end{figure}

\begin{figure} [h]
\begin{center}
\includegraphics[scale=0.65] {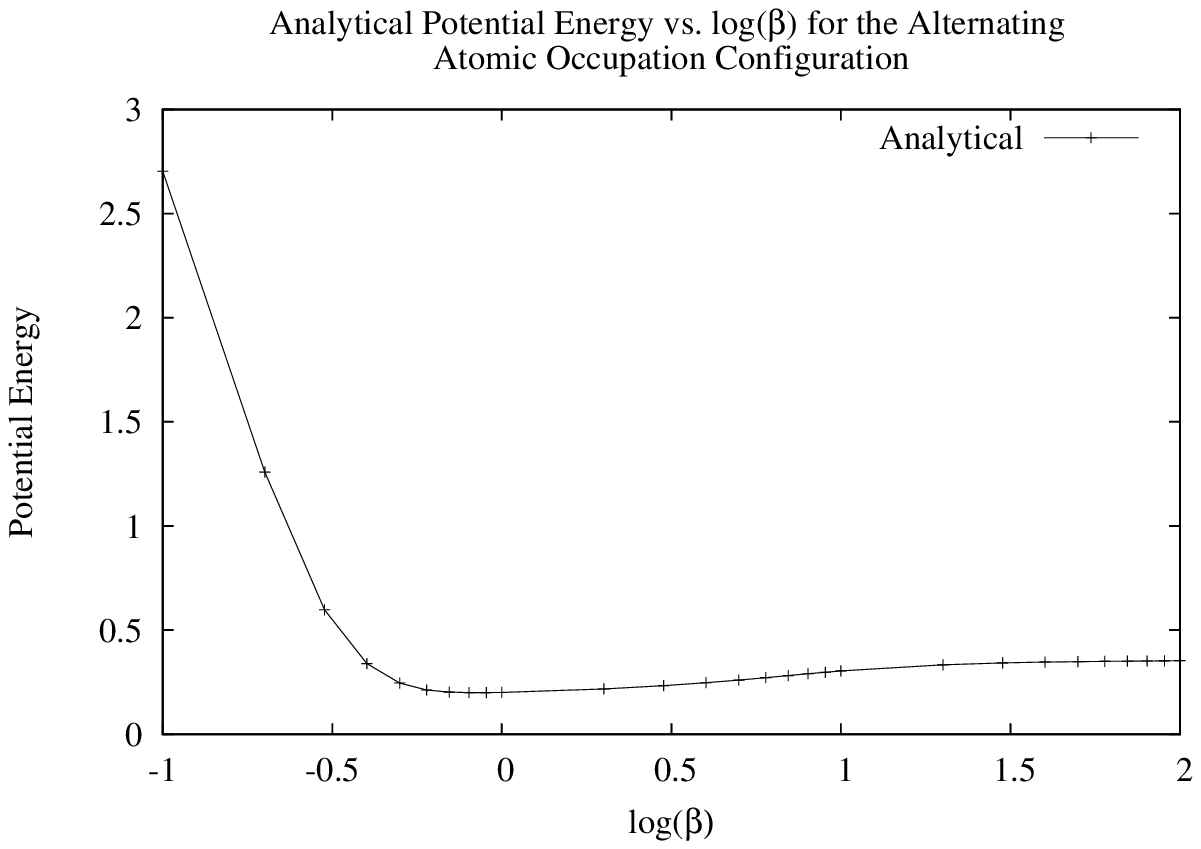} 
\caption{\label {fig:qm/complexfunctions}  The Analytical Potential Energy vs. log($\beta$) for the Striped Case Configuration. The PIMC simulation results are not shown. Note that a larger range of $\beta$ is being considered in order to evaluate asymptotic behavior of the Potential Energy. Due to $\beta$ being considered for a range encompassing four orders of magnitude, log($\beta$) was utilized as the abscissa.}
\end{center}
\end{figure}

{The expression for both kinetic energy operator and the qp-qp correlation function $G_1$ in the classical isomorphism remains the same in the interacting system as that for the free particle, while that for the potential energy is simply given by $V\left(\bf{n}\right) = \sum_{\alpha = 1}^p \epsilon n_{j_{\alpha}}$, where the form of the potential was introduced in Eq. (8). However, in the presence of atoms, we can also define and study the atom-quantum-particle, or atom-qp, correlation function, }
\begin{equation}
G_2\left( n \right) = \langle \sum_j n_j |\psi_{j + n}|^2 \rangle
\end{equation}
{In occupation number representation,}
\begin{equation}
\hat{G}_2 \left( n \right) = \sum_j n_j c_{j + n}^{\dagger} c_{j + n}
\end{equation}
{$G_2$ carries information concerning the range of lattice sites over which the qp wave functions are influenced by an atom, and vice-versa. It is apparent that on an infinite lattice $G_2 \left( n \right)$ vanishes for large $n$ unless the distribution of atoms exhibits long-range order.}

{We now find the path-integral form of this correlation function. As usual, we begin by expressing $G_2\left( n \right)$ as a quantum trace}
\begin{equation}
G_2\left( n \right)  = \frac{Tr\left[ \hat{G_2}e^{-\beta\hat{H'}}\right]}{Z}  \nonumber
\end{equation}
{Following the usual method, we compute the matrix element}
\begin{equation}
\bra{j_1}\hat{G_2}\ket{k} = \bra{j_1}\sum_j n_j c_{j + n}^{\dagger}c_{j + n} \ket{k} \nonumber
\end{equation}
{and we get}
\begin{equation}
\bra{j_1}\hat{G_2}\ket{k} = \bra{j_1} n_j c_{j + n}^{\dagger}\delta_{j + n, k} \ket{\:}    \nonumber
\end{equation}
\begin{equation}
=  \bra{j_1} n_{k - n} c_k^{\dagger}  \ket{\:} \nonumber
\end{equation}
\begin{equation}
=  \bra{j_1} n_{k - n} \ket{k}   \nonumber
\end{equation}
\begin{equation}
= n_{k - n}\braket{j_1}{k}  \nonumber
\end{equation}
\begin{equation}
= n_{k - n} \delta_{j_1, k}  \nonumber
\end{equation}
\begin{equation}
= n_{j_1 - n} \nonumber
\end{equation}
{We next follow the same mathematical steps in computing the traces of other operators previously seen in this paper. In so doing, we obtain}
\begin{equation}
Tr\left[\hat{G_2}e^{-\beta\hat{H'}}\right] = \sum_{j_1} \sum_{j_2}\cdots\sum_{j_p} \left[ \frac{1}{p} \sum_{\alpha = 1}^p n_{j_{\alpha - n}}\right]  \nonumber
\end{equation}
\begin{equation}
\times \prod_{\alpha = 1}^p \bra{j_{\alpha}}e^{-\frac{\beta\hat{H'}}{p}}\ket{j_{\alpha + 1}} \nonumber
\end{equation}
{We now decompose the Hamiltonian in the product into its Kinetic Energy and Potential Energy parts explicitly.}
\begin{equation}
\bra{j_{\alpha}}e^{-\frac{\beta \hat{H'}}{p}} \ket{j_{\alpha + 1}} = \bra{j_{\alpha}}e^{-\frac{\beta \left(\hat{T'} + V_{j_{\alpha}} \right)}{p}} \ket{j_{\alpha + 1}} \nonumber
\end{equation}
{Factoring, we get}
\begin{equation}
\bra{j_{\alpha}}e^{-\frac{\beta \hat{H'}}{p}} \ket{j_{\alpha + 1}} = e^{-\frac{\beta V_{j_{\alpha}}}{p}}\bra{j_{\alpha}}e^{-\frac{\beta \hat{T'}}{p}} \ket{j_{\alpha + 1}} \nonumber  
\end{equation}
{Now, we convert the Kinetic Energy part to the Modified Bessel function equivalent that we have already determined.}
\begin{equation}
\bra{j_{\alpha}}e^{-\frac{\beta \hat{H'}}{p}} \ket{j_{\alpha + 1}} = e^{-\frac{\beta V_{j_{\alpha}}}{p}}I_{j_{\alpha} - j_{\alpha + 1}}\left( \frac{2 \beta t}{p}\right) \nonumber  
\end{equation}
{We can now immediately write the final form of $Tr\left[\hat{G_2}e^{-\beta\hat{H'}}\right]$ and $G_2 \left( n \right)$.}
\begin{equation}
Tr\left[\hat{G_2}e^{-\beta\hat{H'}}\right] = \sum_{j_1} \sum_{j_2}\cdots\sum_{j_p} \left[ \frac{1}{p} \sum_{\alpha = 1}^p n_{j_{\alpha - n}}\right]  \nonumber
\end{equation}
\begin{equation}
\times \prod_{\alpha = 1}^p \left[ e^{-\frac{\beta V_{j_{\alpha}}}{p}}I_{j_{\alpha} - j_{\alpha + 1}}\left( \frac{2 \beta t}{p}\right) \right] \nonumber
\end{equation}
\begin{equation}
G_2 \left( n \right) = \left(\frac{1}{Z}\right)\sum_{j_1} \sum_{j_2}\cdots\sum_{j_p} \left[ \frac{1}{p} \sum_{\alpha = 1}^p n_{j_{\alpha - n}}\right] \nonumber
\end{equation}
\begin{equation}
\times \prod_{\alpha = 1}^p \left[ e^{-\frac{\beta V_{j_{\alpha}}}{p}}I_{j_{\alpha} - j_{\alpha + 1}}\left( \frac{2 \beta t}{p}\right) \right] \Delta\left(\sum_{\alpha = 1}^p s_{\alpha} \right)
\end{equation}
{Thus, $\Gamma_2 \left( n \right)$, the classical isomorphic operator for $G_2 \left( n \right)$, is simply}
\begin{equation}
\Gamma_2 \left( n \right) = \frac{1}{p}\sum_{\alpha = 1}^p n_{j_{\alpha - n}}
\end{equation}
\subsection*{\textit{3. Comparison of analytical calculations and computational results for striped case configuration}}
{Figures 3 through 12 show direct comparisons between analytical calculations performed on Mathematica and the Monte Carlo simulations for the striped configuration. As can be seen in these plots, there was very good agreement between the analytical and computational results. In all cases, there were 100 steps in the random walk, the on-site atomic potential $\epsilon = 10.0$, and every other lattice site of the one-dimensional lattice was occupied by an atom with such a potential. The inverse temperature parameter $\beta$ was varied extensively, from values as small as 0.01 to as large as 100.0 in the appropriate nondimensional units of this paper. }

{Figure 3 is a plot of the Average Potential Energy versus $\beta$ for $\beta$ ranging from 0.0 to 10.0. As can be seen, the analytical and Monte Carlo results matched closely. We see that for $\beta$ = 0.0, the average potential energy is 5.0. This value makes reasonable sense given that for $\beta = 0.0$ there is no penalty for the quantum particle in the random walk to land on an occupied lattice site. So, statistically, at this value of $\beta$ half the quantum particle visits are to occupied sites with on-site potential $\epsilon = 10.0$, and the other half are to unoccupied sites with no on-site potential, giving an overall average potential energy of 5.0. We then see that as $\beta$ increases the analytical and computational potential energy curves quickly tend toward a much lower asymptotic value. One can understand this behavior by looking at Eq. (115) and considering the form of the Gibbs factor. The larger $\beta$ causes the decaying exponential to be smaller regardless. But, for situations where there are several quantum particle positions on occupied sites, the potential summation in the exponential becomes a larger positive number that when multipled by $-\frac{\beta}{p}$ causes the decaying exponential to be even smaller. Hence, any proposed random walk of this nature will likely compute a small acceptance factor $q$. Eventually for large enough $\beta$ the rejection rate will approach 100$\%$. }

{Figures 4 and 5 show the Analytical and Computational Atom-Quantum Particle Correlation Function for the striped configuration, respectively. Notice that these plots are almost identical. In both cases, one notices a general trend that the Atom-Quantum Particle Correlation plots tend to oscillate between a consistent high and low value as the spacing n increases. The larger $\beta$, the greater the size of these oscillations which approach the lower and upper bounds of 0 and 1.  The limiting values are nearly realized for all $\beta \ge 0.5$. Again, there is a penalty for random walks where the qp lands on occupied sites. For large enough $\beta$, the most probable random walk is one where the quantum particle visits only occur on unoccupied sites, which is every other one. Thus, for large enough $\beta$ one sees that the atom-qp correlation function is nearly unity for the vacant sites and nearly zero for the occupied sites. }

{In the low-temperature limit, we see that as we increase $\beta$ there is a point where the equilibrium properties no longer change. For instance, consider Figure 3, the plot of the Average Potential Energy vs. $\beta$ for the striped configuration. By $\beta \approx 0.7$, the $\langle V \rangle$ plot descended to a value of $\sim$0.2 and then the plot gradually rose to an asymptotic value of $\sim$0.35 for higher values of $\beta$. Regarding the Atom-QP correlation plots, Figures 4 and 5, we see that by $\beta = 0.5$ the oscillations in the plots nearly reach bounds of 0 and 1, and this behavior is true for all higher values of $\beta$ considered. Hence, judging from the computational data, it appears that at some value of $\beta$ we are observing ground state dominance. $\left[22\right]$ Recall that Figure 3 shows an analytical and a computational plot for $\langle V \rangle$, where the points on the analytical curve correspond to calculations carried out using Mathematica. Figure 13 shows the plot of the analytically calculated Potential Energy versus log$\left(\beta\right)$. This figure considers the largest range of $\beta$ seen in this paper, where we extended $\beta$ to be as large as 100.0, so that we can best study the asymptotic behavior of this quantity in the low-temperature limit. The abscissa of this plot is logarithmic because $\beta$ varied by four orders of magnitude. One can see from Figure 13 that indeed $\langle V \rangle$ asymptotically approaches a value of $\sim$0.35 for high $\beta\epsilon$ where $\epsilon = 10.0$. Earlier we analytically calculated the ground state potential energy to be about 0.3577, Eq. (98), and this agrees with these analytical calculations in the low-temperature limit.}

{Figures 6 through 12 show the qp-qp correlation function for values of the inverse temperature $\beta$ increasing from 0.01 to 10.0. There is excellent agreement between the analytical and computational results for the qp-qp correlation function for the striped configuration for $\beta = 0.01$ to $\beta = 1.0$, except for some differences in the shoulders due to rare events. These plots are smooth and quickly tend toward 0 asymptotically after a spacing of about $n = 5$. However, for the qp-qp correlation plots corresponding to $\beta = 5.0$ and $\beta = 10.0$, the analytical and computational plots followed each other closely, but there were surprising extra oscillations in the curves, instead of consistently tending toward 0 like the corresponding plots for smaller $\beta$ values.}

{We can qualitatively understand the extra oscillations in Figures 11 and 12 in the following manner. Recall that in the presence of a potential, we calculate the qp-qp correlation in the same manner as that for the free particle, using Eq. (58), where we have a quotient of modified Bessel functions, where the index of the Modified Bessel function in the numerator is $n$ less than the index of the one in the denominator. Hence as $n$ increases, this quotient quickly tends toward zero, since this is how Modified Bessel functions behave. However, due to the Gibbs factor, for large enough $\beta$ the qp is practically restricted to only empty lattice sites, which is every other site for the striped case. Hence, there are two competing influences: 1) there is a tendency for the quantum particle to be concentrated at sites that are multiples of 2 sites away from a given quantum particle location; 2) the free-particle qp-qp correlation function reduces quickly for increasing $n$. These competing influences cause the couple of extra oscillations.}

{It is also interesting to compare a plot of the qp-qp correlation function for the free particle and one for the striped configuration for the same value of inverse temperature $\beta$ to see what is the relative effect of the potential. Figure 2 is a plot of the qp-qp correlation function for the free particle for $\beta$ = 10.0, and Figure 12 is the qp-qp correlation function under the influence of the striped potential for the same value of $\beta$. First of all, the free particle plot showed both the analytical and computational curves descending toward zero at slightly larger spacing $n$ compared with the corresponding striped case plots. Also, the free particle qp-qp correlation plots continually descended toward zero for increased spacing whereas the corresponding striped configuration plots show extra oscillations at larger spacing $n$ before finally descending toward zero monotonically. The effect of the potential seemed to reduce the spread of the quantum particle, but it also caused concentrations of population for discrete spacings of about $n = 3$ and $n = 5$ for larger $\beta$, i.e. $\beta = 5.0$ and $\beta = 10.0$.}
\section*{V. SUMMARY AND CONCLUSIONS}
{In this work we applied path integral Monte Carlo to the case of the extended states of an equilibrated quantum particle on a lattice. The qp experiences the periodic potential resulting from a quenched distribution of atoms. We study the particular case of a low-mass quantum particle interacting with a set configuration of classical atoms on a one-dimensional lattice arranged in an alternating pattern such that every other lattice site is occupied. This configuration produces the most rapid variation in potential, and hence stresses the path integral as much as possible. To be able to perform this study, we first analytically solved the Schrodinger equation for the free particle. We investigated system properties in the canonical ensemble such as the partition function, energy, energy fluctuation and self-correlation of the free quantum particle. Using a path-integral Monte Carlo algorithm developed specifically for this problem of a quantum particle being confined to occupy lattice sites, as opposed to other path-integral algorithms for continuous systems, we established a connection between the quantum trace and the weighted sum of variable-step-sized random walks on the lattice. This isomorphism was used to establish a method for carrying out Monte Carlo calculations of the thermal average of the aforementioned physical observables. As Figures 1 and 2 demonstrate, the agreement between the Monte Carlo results and the analytical calculations for the free quantum particle were within 1$\%$ for the energy but only as good as 20$\%$ for the qp-qp correlation function up to a spacing of $n = 5$. If one considers the qp-qp correlation function beyond $n = 5$ then disagreement occurs because of the occurrence of rare events. }

{Using the same path integral Monte Carlo algorithm with Metropolis sampling and replacing only 20$\%$ of the closed chain per iteration in order to improve the acceptance statistics, the system properties can be solved for a variety of atomic configurations. However, it is only possible to obtain an analytical solution to the Schrodinger equation for very few potentials. Our goal was to construct a non-trivial atomic configuration that could be input into the Monte Carlo code and also be solved analytically. We were able to obtain an analytical solution for the case of a one-dimensional lattice possessing alternating atomic occupation, also known as the striped configuration. We also derived analytical solutions to the partition function, energy, energy fluctuation and the atom-qp correlation function for the striped case, and with the aid of Mathematica we were able to numerically compute results for these analytical formulas. These analytical computations were compared with corresponding Monte Carlo simulations and the agreement for potential energy (Figure 3) was within 1$\%$ for $\beta \le 1.0$ and no worse than 7$\%$ for $\beta \sim 10.0$. The corresponding agreement for the qp-qp correlation functions (Figures 6 - 12) show that the analytical and computation results follow one another, however the error is significantly more than what is observed in the free-particle case. But, the agreement between the analytical and computational results for the atom-qp correlation function (Figures 4 and 5) was within $1\%$ for small $\beta$ and no worse than $5\%$ for $\beta$ = 10.0. Hence, in general the PIMC approach seems to do very well for predictions of energy and atom-qp correlation and only fairly well for qp-qp correlation. Based on the success of the approach demonstrated here, we are planning to employ the path integral method to investigate additional quenched and annealed equilibrium ensembles. In particular, we plan to investigate situations where either Anderson localization or self-trapping of the qp plays the dominant role.}
   
\section*{\large ACKNOWLEDGEMENTS}
{The authors appreciate the support received from the Lockheed Martin Corporation. The authors are grateful to Benjamin Janesko, Jim Mayne, Terrence Reese and Steven Pehrson for their technical assistance.} 
\renewcommand{\refname}

\end{document}